\documentclass[fleqn,usenatbib]{mnras}

\usepackage{newtxtext,newtxmath}

\usepackage[T1]{fontenc}

\DeclareRobustCommand{\VAN}[3]{#2}
\let\VANthebibliography\thebibliography
\def\thebibliography{\DeclareRobustCommand{\VAN}[3]{##3}\VANthebibliography}

\usepackage{graphicx}	
\usepackage{amsmath}	
\usepackage{pdflscape}

\title[Elastic mountain formation via starquakes ]{Applying the starquake model to study the formation of elastic mountains on spinning neutron stars}

\author[Yashaswi Gangwar and David Ian Jones]{
Yashaswi Gangwar \thanks{E-mail: yg2n21@soton.ac.uk}
and David Ian Jones \thanks{E-mail: d.i.jones@soton.ac.uk}
\\
Mathematical Sciences and STAG Research Centre, University of Southampton, Southampton SO17 1BJ, UK\\
}

\date{Accepted XXX. Received YYY; in original form ZZZ}

\pubyear{2023}

\begin{document}
\label{firstpage}
\pagerange{\pageref{firstpage}--\pageref{lastpage}}
\maketitle

\begin{abstract}


When a neutron star is spun-up or spun-down, the changing strains in its solid elastic crust can give rise to sudden fractures known as starquakes.  Early interest in starquakes focused on their possible connection to pulsar glitches.  While modern  glitch models rely on pinned superfluid vorticity rather than crustal fracture,  starquakes may nevertheless play a role in the glitch mechanism.  Recently, there has been interest in the issue of starquakes resulting in non-axisymmetric shape changes, potentially linking the quake phenomenon to the building of neutron star mountains, which would then produce continuous gravitational waves.  Motivated by this issue, we present a simple model that extends the energy minimisation-based calculations, originally developed to model axisymmetric glitches, to also include non-axisymmetric shape changes.  We show that the creation of a mountain in a quake necessarily requires a change in the axisymmetric shape too.  We apply our model to the specific problem of the spin-up of an initially non-rotating star, and estimate the maximum mountain that can be built in such a process, subject only to the constraints of energy and angular momentum conservation.

\end{abstract}




\section{Introduction}

Neutrons stars (NSs) are one of the densest and most compact type of astrophysical object, consisting of 1-2 \(M_\odot\)  in few tens of km of radius making them ideal cosmic laboratories to test physics in extreme conditions (\cite{Blaschke2018}). Since the discovery of NSs in 1967 (\cite{1968Natur.217..709H}), they have been studied using the entire electromagnetic spectrum. (see e.g.\ \cite{Menezes_2021}).

GW170817 was the first observed binary neutron star gravitational wave signal, detected by both LIGO and VIRGO (\cite{abbott2017gw170817}). This detection event started the study of NS physics from a new lens. GW170817 was a transient signal which occurred due to the merger of two NS. We also expect to observe \emph{continuous gravitational waves} (CGWs) signals from spinning non-axisymmetric neutron stars.   No such CGWs have been detected yet, but search efforts are ongoing (see e.g. \citet{Maggiore_2020, 2021arXiv210909882E}). 

A rapidly spinning compact object like a neutron star, when deformed away from an axisymmetric shape, gives rise to a time-varying quadrupole ellipticity (for simplicity we will use the term ``ellipticity" for ``quadrupole ellipticity"). This non-zero ellipticity is termed a ``mountain'', which when spinning produces CGWs (see e.g.\ \citet{Glampedakis2018}). The ellipticity can be quantified as
\begin{equation}
\label{1}
    \epsilon_{22} = \frac{|I_{xx}- I_{yy}|}{I_{zz}},
\end{equation}
where $I_{xx}$, $I_{yy}$ and $I_{zz}$ are the moments of inertia of the star along the $x$, $y$ and $z$-axes respectively, with the rotation occurring along the $z$-axis. The subscript $(22)$ reflects the fact that the corresponding (non-axisymmetric) mass distortion is described by an $(l=2, m=2)$ spherical harmonic. Since the centrifugal forces generated by the rotation gives rise to a symmetrical distortion, corresponding to an $(l=2, m=0)$ perturbation, they do not contribute to the generation of CGWs \citep{Glampedakis2018}.
 
 There are two broad classes of mountain, depending on the physical process which supports these non-axisymmetric deformations. These are: 1) Magnetic mountains: the non-zero ellipticity is supported by Lorentz forces from the magnetic field of the star (see e.g.\ \citet{Glampedakis2018}; \citet{Haskell_Priymak}). 2) Elastic mountains - the non-zero ellipticity is sustained by  elastic strain in the crust of the star (see e.g.\ \cite{10.1111/j.1365-2966.2006.10998.x}, \cite{PhysRevD.88.044004}; \cite{Fattoyev_2018}; \cite{Gittins_2020}; \cite{13}) Thermal mountains come under the broad class of elastic mountains \citep{Bildsten_1998, Ushomirsky_2002, 10.1093/mnras/staa858, hj_23}.

Many spinning neutron stars are observed as radio pulsars.  Their high moments of inertia and steady spin-down torques make them extremely stable clocks. However, observations show occasional sudden increases in their spin rates, termed as \emph{glitches}; see e.g. \cite{11}; \cite{10}. The \emph{starquake} theory was proposed by \cite{Baym_Pines_1971} to explain the glitches observed in the Crab and Vela pulsars, with the glitches being caused by sudden fractures in the star's elastically strained crust.  This model was not able to explain the large glitches observed in Vela (\cite{12}).  The currently preferred model of glitches now involves unpinning of superfluid vortices (\cite{Anderson_Alpar_Pines_Shaham_1981}), but starquakes may nevertheless still occur and may even trigger the unpinning of superfluid vortices to facilitate the glitches (\cite{Epstein:2000iy} ).

In the past, a few attempts were made to study the formation of elastic mountains at glitches using the starquake model.  \cite{Fattoyev_2018} explained the formation of the elastic mountain by considering a portion of the crust moving radially, with the rest of the crust remaining unchanged. As noted by the authors themselves, this is an overly simplistic depiction of the formation of the mountain. We do not expect such radial movement of a part of the crust due to crust break. A more globally consistent model is needed. 

A more quantitative description was given by \cite{13},  who modelled the formation of elastic mountain on an \emph{accreting} NS. They also argued that the centrifugal force acting on the spinning-up NS can be strong enough to break the crust, and that the fracture may itself occur in a non-axisymmetric way.
By making some specific assumptions about the nature of the fracture process, they calculated the range of the starquake-induced ellipticity to be  $10^{-9} - 10^{-5}$.

A detailed study of the non-axisymmetric crust failure on macroscopic scales based on tectonic processes was performed by \cite{10.1093/mnras/stac1351}, who considered a spinning down NS. To model the micro-structure and dynamics of the crust failure, they constructed a cellular automaton. They predicted the rate of crust failure and found that typically the last failure event occurs when the NS spins down to $\approx 1 $ per cent of its birth frequency. They also calculated the ellipticity and gravitational wave strain as a function of the star's age. 

In this paper, we revisit the problem of mountain formation caused by starquakes in spinning-up stars.  We use as our fundamental tool the simple energy minimisation methods first employed by \citet{Baym_Pines_1971}, which allow one to gain intuitive insights into possible NS deformations.  The original \cite{Baym_Pines_1971} analysis specialised to axisymmetric ($m=0$) deformations, as that was the most relevant to the glitch phenomena that they studied.  We extend their analysis to the non-axisymmetric $(m=2$) case, allowing for both axisymmetric and non-axisymmetric deformations simultaneously. 

Our parameter space is two-dimensional (2-d), with one parameter controlling the change in axisymmetric shape of the star, and the other the degree to which it becomes triaxial.  We enforce angular momentum and energy conservation, and examine how large a mountain can be formed, subject only to these constraints.  We also use our model to describe a specific mountain building scenario proposed recently in \citet{13}.

The structure of this paper is as follows.  In Section \ref{section 2}, we briefly explain the starquake model. In Section \ref{section new} we write down a simple geometric model of how the star's shape can change at a starquake.  In Sections \ref{section 3} and \ref{section 4} we give the calculation of the constants related with the strain and perturbed gravitational energy of the star for the ($l=2, m=2$) perturbation, which we need for calculating the mountain size. In Section \ref{section 5}, we perform the calculation to check if the total energy of the star contains cross terms when both symmetrical ($l=2, m=0$) and asymmetrical ($l=2, m=2$) perturbations are present. Section \ref{section 6} gives the relation between the equilibrium shape of the star after the starquake and its corresponding relaxed zero-strain shape of the star for pure $l=2, m=2$ perturbation. In Section \ref{energy_cal} we estimate the change in the total energy of the star during starquake and find the region in the parameter space where the change in the total energy is negative i.e.\ where mountain formation is energetically possible. In Section \ref{2-d space} we use our results to calculate the maximum mountain that can be formed, and make contact with the recent work of \citet{13}.  In Section \ref{sec:summary} we give some concluding remarks.

\section{Energy calculations using the starquake model} \label{section 2}

\begin{figure}
    \centering
    \includegraphics[width=\columnwidth]{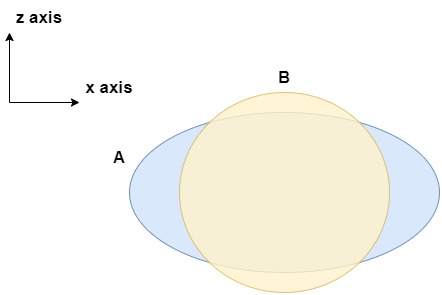}
    \caption{Surface shape for a spinning elastically relaxed star A, that then spins down to give the more slowly spinning elastically strained star B.   Rotation is along the $z$-axis.}    
     \label{fig:fig1}
\end{figure}

\begin{figure}
    \centering
    \includegraphics[width=\columnwidth]{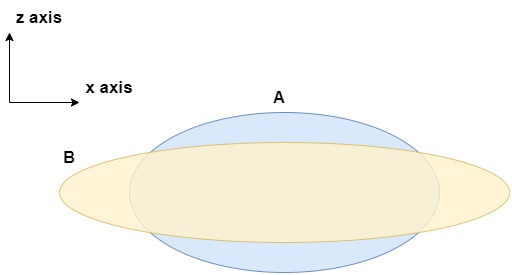}
    \caption{As for Fig. \ref{fig:fig1}, but now the spinning elastically relaxed star A is spun-up to give the more rapidly spinning elastically strained star B.}
    \label{fig:fig2}
\end{figure}

When a star spins up or down, it tries to become more oblate or less oblate respectively due to the change in the centrifugal forces acting on it. This is shown in  Fig. \ref{fig:fig1} and Fig. \ref{fig:fig2}. However, due to the elastic nature of the crust, the star resists this shape change, which results in the development of strain in the crust. This is quantified using the strain tensor
\begin{equation}
\label{equation2}
    \Sigma_{ij} = \frac{1}{2}(\nabla_i \xi_j  + \nabla_j \xi_i),
\end{equation}
 where $\xi_i(\textbf{r})$ is the displacement field connecting the position of the matter elements in the zero-strain NS to the one in the elastically strained configuration. When the strain in the crust reaches a critical limiting value, the crust will break.  This was modelled by \citet{Baym_Pines_1971}, who computed the displacement field $\xi_i$ for a simple uniform density, uniform shear modulus stellar model.  See also \citet{Franco_2000}.

As well as making a detailed calculation of the displacement field, \citet{Baym_Pines_1971} used a simple energy-minimisation model to describe the starquake process, which we will make use of throughout this paper.  In this section we summarise their (axisymmetric) model, which we will later extend to include non-axisymmetry.

Our star is incompressible, of uniform density, and is fluid for radii $0 < r < R_{\rm c}$, and has uniform shear modulus $\mu$ for $R_{\rm c} < r < R$, where $R_{\rm c}$ is the radius of the crust-core boundary, and $R$ is the stellar radius. There are studies which have considered more realistic elastic star with non uniform density and shear modulus as given in \cite{Ushomirsky_2002}, \cite{Cutler:2002np} and \cite{2020MNRAS.491.1064G}. We choose to use an extremely simple model, in order to be able to carry out our calculations analytically. 

The departure of the star's shape from being spherical can be quantified using the \emph{oblateness parameter}  
\begin{equation}
\label{3}
    \epsilon_{20} = \frac{I_{zz} - I_{zz,s}}{I_{zz,s}},
\end{equation}
which is the relative departure of the moment of inertia of the rotating star from the value $I_{zz,s}$ it would have if it were spherical, i.e.\ if it were not rotating and elastically relaxed. The subscript (20) corresponds to the axisymmetric perturbation $l=2, m=0$. 

To find the equilibrium shape one needs to consider the competition between the centrifugal forces caused by rotation, which tend to make the star oblate, the gravitational forces, which tend to make the star less oblate, and elastic forces, that tend to force the star towards the shape that would reduce the elastic strain energy to zero.  The total energy of the star can be written as
\begin{equation}
\label{4}
    E_{\rm{T}} = E_{\rm{grav}} + E_{\rm{rot}} + E_{\rm{strain}} ,
\end{equation}
where $E_{\rm{grav}}$ is the gravitational potential energy, $E_{\rm{rot}}$ is the kinetic energy and $E_{\rm{strain}}$ is the elastic strain energy. We are considering small departures from sphericity.

For the gravitational energy,
\begin{equation}
\label{5}
    E_{\rm{grav}} = E_{\rm{grav},s} + \delta E_{\rm{g}} , 
\end{equation}
where $E_{\rm{grav},s}$ is the gravitational potential energy of the spherical configuration of the star.  The gravitational energy must be a minimum when $\epsilon_{20}=0$, motivating the relation
\begin{equation}
\label{2.5a}
    \delta E_{\rm{g}} = A_{20} \epsilon_{20}^2,
\end{equation}
with $A_{20}$ a constant.  For our uniform density incompressible star of mass $M$ and radius $R$:
\begin{equation}
\label{2.6}
    A_{20}  = \frac{3}{25}\frac{M^2G}{R}.
\end{equation}
This expression for $A_{20}$ is mentioned in \cite{Baym_Pines_1971}. They did not derive this expression, but instead gave the reference \citet{love2013treatise}.  Since we did not find the derivation in \cite{love2013treatise}, we derived the expression for $A_{20}$ explicitly using the method set out in Section 7.3 of \citet{SHAPIRO}; see also our Appendix \ref{App2}.

For the kinetic energy,
\begin{equation}
\label{9}
    E_{\rm{rot}} =\frac{L^2}{2I_{zz}},
\end{equation}
where $L$ is the angular momentum.  

The moment of inertia about the rotation axis, $I_{zz}$, can be written in terms of the oblateness parameter using equation (\ref{3}):
\begin{equation}
\label{n18}
    I_{zz} = I_{zz,s}(1+\epsilon_{20}). 
\end{equation}

The elastic strain energy must be minimised when the actual shape $\epsilon_{20}$ is equal to the 
shape $\epsilon_{20,0}$  at which the strain is zero, motivating the relation
\begin{equation}
\label{7}
    E_{\rm{strain}} = B_{20}(\epsilon_{20} - \epsilon_{20,0})^2,
\end{equation}
where $B_{20}$ is a constant.  For our uniform shear modulus crust, with inner and outer radii $R_c$ and $R$, 
\begin{equation}
\label{8}
    B_{20} = \frac{57}{50} \mu  V_{\rm{c}} ,
\end{equation}
where $V_{\rm{c}}$ is the volume of the crust \citep{Baym_Pines_1971}. We have replaced the total volume V in the expression of $B_{20}$, as given in \cite{Baym_Pines_1971}, with the volume of the crust $V_{\rm{c}}$. We can write  expression (\ref{8}) in terms of the thickness of the crust $\Delta R$ as,
\begin{equation}
\label{B20_scaled}
    B_{20} = \frac{57}{50}  (4\pi R^2 \Delta R ) \mu.
\end{equation}

Collecting the above pieces the total energy is
\begin{equation}
\label{10}
    E_{\rm{T}} = E_{\rm{grav},s} + \frac{L^2}{2I_{zz,s}(1+\epsilon_{20})} +   A_{20}\epsilon_{20}^2
    + B_{20}(\epsilon_{20} - \epsilon_{20,0})^2 .
\end{equation}

To calculate the equilibrium shape $\epsilon_{20}$, minimize $E_{\rm{T}}$ with respect to $\epsilon_{20}$, keeping the angular momentum $L$ constant:
\begin{equation}
\label{11}
    \epsilon_{20} = \frac{I_{zz,s} \Omega^2 }{4(A_{20}+B_{20})} + \frac{B_{20}}{A_{20}+B_{20}}\epsilon_{20,0} .
\end{equation}
This equation gives the relation between the equilibrium shape ($\epsilon_{20}$) and the elastically relaxed shape ($\epsilon_{20,0}$) of the star. The first term, proportional to $\Omega^2$, can be identified as the centrifugal distortion.  The second term, which remains when $\Omega$ is set to zero, is the (axisymmetric) distortion supported by the elastic strains in the crust. 

We can write equation (\ref{11}) as,
\begin{equation}
    \epsilon_{20} = \epsilon_{\Omega} + b_{20}\epsilon_{20,0},
\end{equation}
where
\begin{equation}
\label{ep_omega}
    \epsilon_{\Omega} = \frac{5}{6}\frac{R^3 \Omega^2}{G M} = 1.76 \times 10^{-3} \left  (\frac{f}{100 \hspace{0.1 cm} \rm{Hz}} \right)^2\frac{R_6^3}{M_{1.4}},
\end{equation}
using $\Omega = 2  \pi f$, and
\begin{multline}
\label{b20_scaled}
    b_{20} = \frac{B_{20}}{A_{20}+B_{20}} \approx \frac{B_{20}}{A_{20}} = \frac{38 \pi  \mu   R^3 \Delta R}{M^2 G} \approx 2.28 \times 10^{-5} \frac{ \mu_{30} R_6^3 \Delta R_5  }{M_{1.4}^2} .
\end{multline}
In (\ref{ep_omega}) and (\ref{b20_scaled}), we have scaled to the canonical values, $M_{1.4}$ the mass in units of $1.4 M_\odot$, $R_6$ the NS radius in units of  $10^6 \rm{cm} $, $\mu$ the shear modulus of the crust in units of $10^{30}  \hspace{0.1 cm} \rm{erg} \hspace{0.1 cm} \rm{cm^{-3}}$ and $\Delta R_5$ the thickness of the crust in units of $10^5 \hspace{0.1 cm} \rm{cm}$.

This analysis is not applicable to mountains as it is specific to axisymmetric perturbations, but can be modified to consider $l=2, m=2$ perturbations. If we neglect rotation, we
can straight-forwardly modify equation (\ref{10}) to give
\begin{equation}
\label{eq:E_T_m=2}
    E_{\rm{T}} = E_{\rm{grav},s}  +   A_{22}\epsilon_{22}^2 + B_{22}(\epsilon_{22} - \epsilon_{22,0})^2 ,
\end{equation}
where the parameters $\epsilon_{22}$ and $\epsilon_{22,0}$ describe the shape of the actual equilibrium configuration, and the zero-strain configuration, both non-axisymmetric.  Minimisation with respect to $\epsilon_{22}$ then gives
\begin{equation}
\label{11c}
   \epsilon_{22}  = \frac{B_{22}}{A_{22}+B_{22}}\epsilon_{22,0}.
\end{equation}
This gives the relation between the equilibrium shape ($\epsilon_{22}$) and the elastically relaxed shape ($\epsilon_{22,0}$) of the star for the non-axisymmetric ($l=2, m=2$) perturbation. 

In the literature till now, the values of $A_{22}$ and $B_{22}$ were assumed to be identical to $A_{20}$ and $B_{20}$. Since, we have $A_{20} \gg B_{20}$, we can make the approximation
\begin{equation}
\label{11d}
   \epsilon_{22}  = \frac{B_{22}}{A_{22}+B_{22}}\epsilon_{22,0} \approx \frac{B_{22}}{A_{22}}\epsilon_{22,0}
   \approx \frac{B_{20}}{A_{20}}\epsilon_{22,0}
   \approx 10^{-5}\epsilon_{22,0},
\end{equation}
where the last equality is motivated by equation (\ref{b20_scaled}).   This indicates that the actual mountain size is always much smaller than its relaxed $m=2$ shape. In Sections \ref{section 3} and \ref{section 4} we will explicitly calculate the values of $B_{22}$ and $A_{22}$ and verify that this is indeed the case.

\section{Displacement vector field and ellipticity}
\label{section new}

In order to model triaxial deformations, we will need the form of the displacements field $\vec \xi$ that generates the non-axisymmetric perturbation from the spherical configuration. This immediately presents us with a problem.  In the axisymmetric perturbation ($l=2, m=0$) we had rotation as the physical deforming process, which enabled computation of a unique displacement field in the strained star (\cite{Baym_Pines_1971}; \cite{Keer2014DevelopingAM}). In the non-axisymmetric perturbation ($l=2, m=2$), we do not have such an \emph{a priori} physical deforming process.  We will therefore have to choose for ourselves the form of a 1-parameter family of solutions. 
  
In our calculation we  choose the following displacement field:
\begin{equation}
\label{m2_displacment_field}
     \Vec{\xi_{22}} = \alpha_{22}\nabla(r^2 Y_{22}(\theta, \phi)).
\end{equation}
Strictly, one has to take the real part of the displacement field vector $\Vec{\xi_{22}}$ to obtain the physical displacement.  The parameter $\alpha_{22}$ controls the size of the perturbation. This is in fact precisely the form of the displacement field one gets when computing the $l=2, m=2$ \emph{oscillation mode} of a fluid, incompressible, uniform density non-rotating spherical star. These \emph{oscillation modes} ($\Vec{\xi_{lm}}$) are called \emph{Kelvin modes} \citep{thomson1863xxviii}).  For this reason we will term this the ``Kelvin mode'' displacement field, but it should be understood that our displacements are constant in time (aside from the time generation generated by the rigid rotation of the star), not oscillatory.

For reasons of simplicity, we will continue to follow the treatment of \citet{Baym_Pines_1971} and model our NSs as incompressible, uniform density, with a non-zero (but uniform) shear modulus $\mu$ only for the crustal region $R_{\rm c} < r < R$.  Since the energies involved in our calculations are of the second order (in the parameter $\alpha_{22}$), we need to verify if the displacement vector field we have chosen ensures the incompressibility of the star to the second order. However, we immediately find a problem---the displacement field of equation (\ref{m2_displacment_field}) gives a \emph{contraction} of the star if one computes to second order in $\alpha_{22}$.  This issue of lack of volume conservation to second order was first pointed out in \cite{Yim_2022} who modelled Kelvin modes proper (i.e.\ stellar oscillations),  who noted that it is a consequence of the  Kelvin mode solutions being calculated to only first order. 

This problem can be most easily demonstrated by calculating the volume change due to the perturbation.  If we allow the displacement field to act on a spherical star of radius $R$, volume $V = 4 \pi R^3 /3$, the deformed star is a (triaxial) ellipsoid, with radii along $x$, $y$ and $z$ axes:
\begin{equation}
\label{106}
    a_1 = R+ \xi^r(r=R, \theta = \frac{\pi}{2} , \phi = 0) = R(1+     \alpha_{22} \frac{1}{2}\sqrt{\frac{15}{2\pi}} ) ,
\end{equation}
\begin{equation}
\label{107}
    a_2 = R+ \xi^r(r=R, \theta = \frac{\pi}{2} , \phi = \frac{\pi}{2}) = R(1 -  \alpha_{22} \frac{1}{2}\sqrt{\frac{15}{2\pi}} ) ,
\end{equation}
\begin{equation}
\label{108}
    a_3 = R+ \xi^r(r=R, \theta = 0 ) = R,
\end{equation}
where  $ \xi^r$ is the radial component of the  displacement field. As the deformed star is an ellipsoid, its volume is given by the standard formula 
\begin{equation}
\label{109}
    V = \frac{4 \pi}{3}a_1 a_2 a_3.
\end{equation}
If $V_{\rm S}$ denotes the volume of the spherical star of radius $R$, the fractional change in the volume is then
\begin{equation}
\label{110}
    \frac{V - V_{\rm S}}{V_{\rm S}} = -  \frac{15}{8\pi} \alpha^2_{22},
\end{equation}
demonstrating the reduction in volume.

To gain more insight, we can compute the corresponding (Lagrangian) density perturbation using equation (B32) from \citet{FRIEDMAN}
\begin{equation}
    \frac{\Delta \rho}{\rho} = - \nabla_i \xi^i +\frac{1}{2}(\nabla_i \xi^i \nabla_j \xi^j + \nabla_i \xi^j \nabla_j \xi^i) + \mathcal{O}(\xi^3).
\end{equation}
We obtain a non-zero value for the second order density perturbation, independent of position within the star:
\begin{equation}
    \frac{\Delta \rho}{\rho} = \frac{15}{8\pi} \alpha^2_{22}.
\end{equation}
We see that the decrease in the volume is compensated exactly by the increase in the density, consistent with conservation of mass.

This shrinking of the star is reflected in a non-zero value for the change in the \emph{internal energy} of the NS, something which we would expect to be zero for an incompressible star, and therefore neglected in the formulae of Section \ref{section 2}.  The change in internal energy $\delta E_{\rm{int}}$ can be computed using equation (B48) of \citet{FRIEDMAN}, which gives
\begin{equation}
     \delta E_{\rm{int}} = \frac{3}{8} \frac{GM^2}{\pi R}\alpha^2_{22}.
\end{equation}
The shrinking of the star has resulted in a positive perturbed internal energy, of second order in $\alpha_{22}$.

The spatial uniformity of the density perturbation indicated that the Kelvin mode displacement field is producing a contraction linear in radius $r$. To enforce incompressibility to second order we therefore add a corresponding spherical expansion to compensate:
\begin{equation}
\label{111}
     \Vec{\xi_{22}} = \alpha_{22}\nabla(r^2 Y_{22}(\theta, \phi)) + \frac{5r}{8\pi}\alpha^2_{22}   \hat{e_r} .
\end{equation} 
We can interpret this as, along with the $Y_{22}$ perturbation there is also a $Y_{00}$ perturbation which causes the spherically symmetric expansion of the radius of the star. The radii of the star along $x$, $y$ and $z$-axes becomes 
\begin{equation}
\label{112}
    a_1 =  R+\xi^r(r=R, \theta = \frac{\pi}{2} , \phi = 0) = R(1+ \frac{5}{8\pi}\alpha^2_{22})(1+     \alpha_{22} \frac{1}{2}\sqrt{\frac{15}{2\pi}} ) ,
\end{equation}
\begin{equation}
\label{113}
    a_2 =  R+\xi^r(r=R, \theta = \frac{\pi}{2} , \phi = \frac{\pi}{2}) = R(1+ \frac{5}{8\pi}\alpha^2_{22})(1 -  \alpha_{22} \frac{1}{2}\sqrt{\frac{15}{2\pi}} ) ,
\end{equation}
\begin{equation}
\label{114}
    a_3 = R+ \xi^r(r=R, \theta = 0 ) = R(1+ \frac{5}{8\pi}\alpha^2_{22}).
\end{equation}
One can easily verify that this modification gives zero density, volume and internal energy  perturbations up to second order in $\alpha_{22}$.  In what follows, we will therefore take equation (\ref{111}) as our 1-parameter family of ($l=2, m=2$) stellar deformations.

Note that for the uniform density triaxial ellipsoids that we consider in this paper, once one has specified the radii $a_1, a_2, a_3$, one can immediately compute the corresponding moments of inertia $I_{xx}, I_{yy}, I_{zz}$ using the (exact) formulae \citep{chandrasekhar1969ellipsoidal}
\begin{eqnarray}
\label{I_xx}
    I_{xx} = \frac{M (a_2^2 + a_3^2)}{5}, \\
\label{I_yy}
    I_{yy} = \frac{M (a_3^2 + a_1^2)}{5}, \\
\label{I_zz}
    I_{zz} = \frac{M (a_1^2 + a_2^2)}{5}.
\end{eqnarray}
Substituting these into the definition of $\epsilon_{22}$ of equation (\ref{1}) gives
\begin{equation}
\label{eq:epsilon_22_radii}
    \epsilon_{22} = \frac{a_1^2 - a_2^2}{a_1^2 + a_2^2}
\end{equation}
Then substituting for the radii $a_1, a_2$ using equations (\ref{112}) and (\ref{113}) gives, to leading order in $\alpha_{22}$,
\begin{equation}
\label{eq:epsilon_22-alpha_22}
    \epsilon_{22} = \sqrt{\frac{15}{2\pi}} \alpha_{22}.
\end{equation}
This is useful as it will allow us to convert between ellipticities and vector field displacements.

\section{Calculation of the strain energy}
\label{section 3}
 
In this section we will calculate the strain energy of the triaxially-deformed star and the corresponding constant $B_{22}$, i.e.\ find the $m=2$ version of equation (\ref{8}).  In Section \ref{section 4} we will consider the (perturbed) gravitational energy, computing $A_{22}$, i.e.\ the $m=2$ version of equation (\ref{2.6}).  

\begin{figure}
    \centering
    \includegraphics[width=\columnwidth]{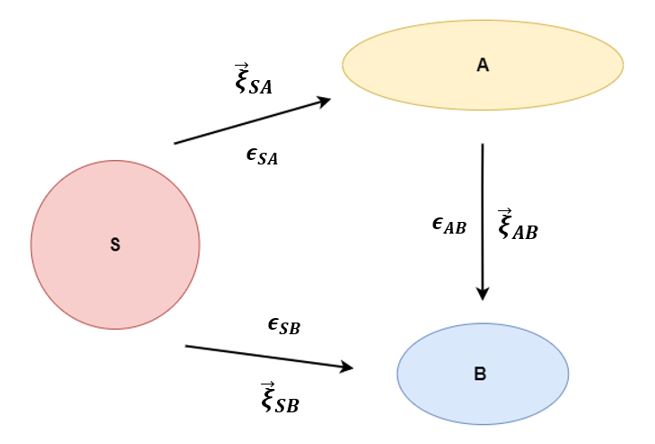}
    \caption{Schematic picture of the three stars relevant to the computation of the constants $A_{22}$ and $B_{22}$.  Star S is the background spherical star. A is the non-spherical zero-strain configuration and B is the elastically strained equilibrium configuration of the star. $\Vec{\xi_{SA}}$ and $\Vec{\xi_{SB}}$ are the displacement fields of the star A and B w.r.t. to the background spherical star S. $\Vec{\xi_{AB}}$ is the displacement field of the star B w.r.t. to its zero-strain configuration A. Similarly, $\epsilon_{SA}$ and $\epsilon_{SB}$ are the ellipticities of the star A and B w.r.t. to the background spherical star ($\epsilon_{S} =0$). $\epsilon_{AB}$ is the difference in ellipticity of the star A and B.}
     \label{fig:7}
\end{figure}

In performing these calculations, it is useful to identify several related stellar configurations, and the displacement fields that connect them, as summarised in Fig. \ref{fig:7}.  Star S is non-rotating, elastically relaxed, and therefore spherical. Star A is non-spherical but elastically relaxed, i.e. has zero strain; it is not an equilibrium solution.  Star B is the actual equilibrium solution, where strains in the now-stressed elastic crust contribute to the force balance.  It is the form of star B that we ultimately wish to calculate.   All three stars have the same mass.  $\Vec{\xi_{SA}}$ and $\Vec{\xi_{SB}}$ are the displacement fields of A and B w.r.t. to the background spherical star S. $\Vec{\xi_{AB}}$ is the displacement field of the star B w.r.t. to  its zero-strain configuration A. Similarly, $\epsilon_{SA}$ and $\epsilon_{SB}$ are the ellipticities of A and B w.r.t. to the background spherical star ($\epsilon_{S} =0$). $\epsilon_{AB}$ is the difference in ellipticity of stars A and B.

As the neutron star spins up (or down), it develops strain in its crust. This gives rise to the strain energy (\cite{thorne2017modern}, Section: 11.4)
\begin{equation}
\label{b1}
    E_{\rm{strain}} = \int U dV 
\end{equation}
where $U$ is the strain energy density given as
\begin{equation}
\label{b2}
   U  = \mu \Sigma_{ij}\Sigma_{ij}.
\end{equation}
where $\mu$ is the uniform shear modulus and $\Sigma_{ij}$ the strain tensor of equation (\ref{equation2}) 
\begin{equation}
\label{b3}
    \Sigma_{ij} = \frac{1}{2}(\xi^{AB}_{i;j} + \xi^{AB}_{j;i}),
\end{equation}
where we use the Kelvin mode displacement field vector $\Vec{\xi}^{AB}$ (\ref{111}) for the $(l=2, m=2)$ perturbation which connects the fluid elements between star A and star B as shown in Fig. \ref{fig:7}. 

Next, we calculate the different components of the strain tensor. Expressions for the complete strain tensor $\Sigma_{ij}$ are given in Appendix A. Inserting (\ref{b18}) into (\ref{b2}), we get
\begin{equation}
    U = \mu \Sigma_{ij} \Sigma_{ij} = \frac{15}{4\pi}\mu (\alpha^{AB}_{22})^2 ,
\end{equation}
where $\alpha_{22}^{AB}$ is the amplitude of the displacement field $\vec \xi^{AB}$ of equation (\ref{111}).
We see that the  strain energy density is uniform, i.e.\ independent of position. The volume integration of the strain energy density $U$ over the crust of the star is then a simple multiplication by the crustal volume (the spherical shell $R_{\rm c} < r < R$), giving the total strain energy:
\begin{equation}
\label{b19}
    E_{\rm{strain}} = 5 (R^3 - R_c^3)\mu (\alpha^{AB}_{22})^2.
\end{equation}

The phenomenological expression for the strain energy in terms of ellipticity is given as per equation  (\ref{7}) above,
\begin{equation}
\label{20}
 E_{\rm{strain}} = B_{22} (\epsilon^{SB}_{22} - \epsilon^{SA}_{22,0})^2,
\end{equation}
where $\epsilon^{SB}_{22}$ is the ellipticity of the equilibrium configuration B of the star as shown in Fig. \ref{fig:7}, and $\epsilon^{SA}_{22,0}$ is the ellipticity of star A.  By definition, these two stars are connected by the displacement field $\xi_{AB}$ and so, using equation (\ref{eq:epsilon_22-alpha_22}), we have
\begin{equation}
\label{b40}
     \epsilon^{SB}_{22} - \epsilon^{SA}_{22,0}  
     = \epsilon^{AB}_{22}  = \sqrt{\frac{15}{2\pi} }\alpha^{AB}_{22} .
\end{equation}

Inserting (\ref{b40}) into the phenomenological expression of the strain energy (\ref{20}) and then comparing it with (\ref{b19}) we get,
\begin{equation}
\label{97}
    B_{22} = \frac{1}{2} V_{\rm{c}} \mu ,
\end{equation}
where $V_{\rm{c}}$ is the volume of the crust.  Comparing (\ref{8}) and (\ref{97}), we get the following relation between $B_{20}$ and $B_{22}$,
\begin{equation}
\label{b20_b22}
    B_{22} = \frac{25}{57} B_{20} .
\end{equation}
We will use this result later.

\section{Calculating the gravitational potential energy }
\label{section 4}

In this section we will calculate the perturbed gravitational energy for the case of $l=2, m=2$, using the formalism described in \citet{FRIEDMAN}, i.e.\ find the $m=2$ version of equation (\ref{2.6}).  As a check on our result, we also obtain the gravitational energy perturbation using formalisms described in  \cite{SHAPIRO} and in \citet{chandrasekhar1969ellipsoidal}. Calculation under \cite{SHAPIRO} formalism is shown in Appendix \ref{App2}.

Equation (B56) of  \citet{FRIEDMAN} give an expression for perturbed gravitational energy accurate to second order in the displacement field $\xi$:
\begin{multline}
\label{98}
    \delta E_{\rm{grav}} = \int_V \left[ \rho \xi^i \nabla_i \Phi + \rho \xi^i \nabla_i \delta \Phi  + \frac{1}{8 \pi G} \nabla_i \delta \Phi \nabla^i \delta \Phi \right.  \\ \left. \quad + \frac{1}{2}\rho \xi^i \xi^j \nabla_i \nabla_j \Phi \right] dV,
\end{multline} 
where  $\Phi$ is the unperturbed gravitational potential inside the star.  

For our uniform density star, the unperturbed $r<R$ gravitational potential is given by
\begin{equation}
\label{99}
    \Phi = \frac{2}{3}\pi G \rho (r^2 -3R^2).
\end{equation}

The relevant perturbation is the Kelvin mode displacement field vector $\Vec{\xi}^{SB}$  which connects the fluid elements between star S and star B as shown in Fig. \ref{fig:7}, as per equation (\ref{111}).    We also need the corresponding perturbed gravitational potential, $\delta \Phi$. 

An important point to notice here is the domain of integration in equation (\ref{98}) is infinite. Therefore, while calculating the third term of (\ref{98}), one has to consider both $\delta\Phi_{\rm{int}} (0 \leq r \leq R)$  and $\delta\Phi_{\rm{ext}} ( r > R )$.  These are given in  \citet{Yim_2022}:
\begin{equation}
\label{100}
    \delta\Phi_{\rm{int}} = -\frac{4 \pi \rho G l}{2l+1} \alpha_{lm} \frac{r^l}{R^{l-2}} Y_{lm}(\theta, \phi),
\end{equation}
\begin{equation}
\label{101}
    \delta\Phi_{\rm{ext}} = -\frac{4 \pi \rho G l}{2l+1} \alpha_{lm} \frac{R^{l+3}}{r^{l+1}} Y_{lm}(\theta, \phi).
\end{equation}
Note that the crustal shear modulus does not appear in either of these expressions for the perturbed gravitational potential; the potential energy depends only the shape of the star, not its composition.

Inserting all the required terms into (\ref{98}) and performing the integration, we obtain
\begin{equation}
\label{115}
    \delta E_{\rm{grav}}  = \frac{3}{10 \pi R} G M^2 (\alpha^{SB}_{22})^2.
\end{equation}

To calculate the constant $A_{22}$ we will equate (\ref{115}) with the phenomenological expression for the perturbed gravitational energy 
\begin{equation}
\label{116}
    \delta E_{\rm{grav}}  = A_{22}(\epsilon^{SB}_{22})^2,
\end{equation}
where
\begin{equation}
\label{b20_again}
    \epsilon^{SB}_{22} = \frac{|\Delta^{SB} I_{xx}-\Delta^{SB} I_{yy}|}{I_{zz,s}}.
\end{equation}
Here $\Delta^{SB} I_{xx}$ and $\Delta^{SB} I_{yy}$ are the differences in moment of inertia of the triaxial strained star B along the x and y axis respectively w.r.t. to the spherical configuration S. The eccentricity parameter $\epsilon^{SB}_{22}$ is related to the Kelvin mode parameter $\alpha^{SB}_{22}$ in exactly the same way as the parameter $\epsilon^{AB}_{22}$ is related to the parameter $\alpha^{AB}_{22}$ as given in equation (\ref{eq:epsilon_22-alpha_22}) above, i.e.
\begin{equation}
\label{116a}
\epsilon^{SB}_{22} = \sqrt{\frac{15}{2\pi} }\alpha^{SB}_{22}. 
\end{equation}
Comparing (\ref{115}) with (\ref{116}) we get,
\begin{equation}
\label{3.146g}
    \frac{3}{10 \pi R} G M^2 (\alpha^{SB}_{22})^2 = A_{22} (\epsilon^{SB}_{22})^2.
\end{equation}
Inserting the expression of  ellipticity $\epsilon^{SB}_{22}$ (\ref{116a}) into (\ref{3.146g}) we get,
\begin{equation}
\label{a22}
    A_{22}= \frac{ GM^2}{25 R}.
\end{equation}
This gives us the value of the constant $A_{22}$. Taking the ratio of $A_{20}$ (\ref{2.6}) and $A_{22}$ (\ref{a22}), we get the following relation,
\begin{equation}
\label{a20_a22}
A_{22} = \frac{1}{3} A_{20} . 
\end{equation}
We will use this result later.

\section{Allowing for both axisymmetric and non-axisymmetric perturbations}
\label{section 5}

In a real NS, there will be axisymmetric strains, due to spin-up/spin-down, and also non-axisymmetric ones, supporting a mountain.  We therefore need to model both $(l=2, m=0)$ and $(l=2, m=2)$ deformations \emph{simultaneously}.  The expression for the energy of such a star will certainly contain the terms given in equations (\ref{10}) and (\ref{eq:E_T_m=2}) for the separate contributions, but one can ask, will there be cross terms, i.e. terms proportional to the product of the small parameters describing each sort of perturbation?  In this section we show there are no such cross terms.

In the case of slow rotation of uniform density stars, the form of the $l=2, m=0$ perturbation are known, as described in \citet{Baym_Pines_1971} and \citet{Keer2014DevelopingAM})
\begin{equation}
\label{d1}
    \Vec{\xi_{20}} = U(r)Y_{20}(\theta) \hat{r} + \frac{V(r)}{r}\frac{d Y_{20}(\theta)}{d \theta}\hat{\theta}
\end{equation}
where
\begin{equation}
    U(r) =\frac{2}{5R^2}\sqrt{\frac{\pi}{5}} \frac{\lambda}{1+b_{20}}(\Omega_B^2 - \Omega_A^2) (3r^3 - 8R^2 r)
\end{equation}
 and 
\begin{equation}
    V(r) =\frac{2}{5R^2}\sqrt{\frac{\pi}{5}} \frac{\lambda}{1+b_{20}}(\Omega_B^2 - \Omega_A^2)(\frac{5}{2}r^4- 4R^2r^2).
\end{equation}
Here, $\lambda = \frac{5}{8\pi G \rho}$.  Note that expression (\ref{d1}) is strictly true only for a uniform density elastic star (i.e. the whole star is elastic), for which \cite{Baym_Pines_1971}  find $b_{20} = \frac{57\mu}{8\pi G \rho^2 R^2}$. We make the approximation that (\ref{d1}) is the correct eigenfunction for the star having finite crust thickness $\Delta R$ with $b_{20}$ as given in equation (\ref{b20_scaled}).  This should be a good approximation given the small effect of elasticity on the shape of the star. 

The displacement field (\ref{d1}) relates two configurations of the star, which can be identified with stars A and B of Fig. \ref{fig:7}, if we now picture both stars as axisymmetric, with A being relaxed and having an angular velocity $\Omega_A$, and B being obtained by spinning-up/down A to angular velocity $\Omega_B$, and therefore being strained.  As before, star S is a reference star of the same mass, but spherical and unstrained.  Then we can take
\begin{equation}
\label{d2}
    \alpha_{20} = \frac{\lambda}{1+b_{20}}(\Omega_B^2 - \Omega_A^2),
\end{equation}
as the small dimensionless parameter describing the effect of rotation on the star. 
 
Given (\ref{d1}), we can calculate the radii along the $x$, $y$ and $z$ axes for any uniform density spherical elastic star that is spin-up from $\Omega_A = 0$ to $\Omega_B > 0$:
\begin{equation}
\label{a_1_elastic}
    a_1 = R+ \xi^r(r=R, \theta = \frac{\pi}{2} , \phi = 0) = R(1+     \alpha_{20} \frac{1}{2}) ,
\end{equation}
\begin{equation}
\label{a_2_elastic}
    a_2 = R+ \xi^r(r=R, \theta = \frac{\pi}{2} , \phi = \frac{\pi}{2}) = R(1 +  \alpha_{20} \frac{1}{2} ) ,
\end{equation}
\begin{equation}
\label{a_3_elastic}
    a_3 = R+ \xi^r(r=R, \theta = 0 ) = R(1 - \alpha_{20} ),
\end{equation}
where  $ \xi^r$ is the radial component of the  displacement field (\ref{d1}). 

We can then calculate the corresponding oblateness parameter $\epsilon_{20}$, as defined in equation (\ref{3}).
First, using equation (\ref{I_zz}) in equation (\ref{3}) we have
\begin{equation}
\label{eq:epsilon_20_radii}
    \epsilon_{20} =  \frac{( a_1^2 + a_2^2 ) - 2R^2}{2R^2}.
\end{equation}
Then, substituting the radii $a_1$ and $a_2$ of equations (\ref{a_1_elastic}) and (\ref{a_2_elastic}) we obtain
\begin{equation}
\label{ep_alpha}
    \epsilon_{20} = \alpha_{20} + \frac{\alpha^2_{20}}{4}.
\end{equation}
For $\alpha_{20} \ll 1$, this can be inverted:
\begin{equation}
\label{alpha_ep}
    \alpha_{20} \approx \epsilon_{20 } -  \frac{\epsilon^2_{20}}{4}.
\end{equation}
This result will prove useful, as it will allow us to translate between oblateness and vector field displacements.

When both the perturbations  $l=2 ,m=0$ and $l=2 , m=2$ are present, we can write the full displacement field as the linear combination of $\Vec{\xi_{20}}$ of equation (\ref{d1}) for and $\Vec{\xi_{22}}$ of equation (\ref{111}), so that
\begin{equation}
\label{118}
     \Vec{\xi} = \alpha_{20} \Vec{\xi_{20}} + \alpha_{22}\Vec{\xi_{22}}. 
\end{equation}
The parameter $\alpha_{22}$, as discussed in Section \ref{section new}, controls the size of the $l=2, m=2$ perturbation. The expression for the total energy of the spinning star can then be written as
\begin{multline}
   \label{117}
    E_{\rm{total}} = E_{\rm{grav,s}} + \frac{L^2}{2I_{zz}} + A_{20}(\epsilon^{SB}_{20})^2 +B_{20}(\epsilon^{SB}_{20} -\epsilon^{SA}_{20,0})^2  + A_{22}(\epsilon^{SB}_{22})^2 + \\B_{22}(\epsilon^{SB}_{22} -\epsilon^{SA}_{22,0})^2    + A_{20,22}\epsilon^{SB}_{20}\epsilon^{SB}_{22} + B_{20,22}(\epsilon^{SB}_{20} -\epsilon^{SA}_{20,0})(\epsilon^{SB}_{22} -\epsilon^{SA}_{22,0}), 
\end{multline}
where $\epsilon^{SB}_{20}$ and $\epsilon^{SB}_{22}$ are the equilibrium oblateness and ellipticity as mentioned in the previous sections. $\epsilon^{SA}_{20,0}$ and $\epsilon^{SA}_{22,0}$ are the zero-strain configurations which can be mapped from the background spherical configuration using the rotation (\ref{d1}) and  Kelvin mode displacement field $\Vec{\xi}$ (\ref{111}) respectively. 
Note that we have allowed for the presence of cross-terms in both the gravitational potential energy and in the elastic energy through the introduction of new parameters $A_{20, 22}$ and $B_{20, 22}$ respectively.  

First consider the elastic strain energy. The methodology used here is similar to the one we followed for the  calculation of $B_{22}$ in Section \ref{section 3}. For the displacement field (\ref{118}), we get
\begin{multline}
\label{3.229}
     \Sigma_{ij} \Sigma_{ij} = (\alpha^{AB}_{20})^2 (\Sigma_{ij}^{20})^2 + (\alpha^{AB}_{22})^2 (\Sigma_{ij}^{22})^2 + \alpha^{AB}_{20} \alpha^{AB}_{22}  \Sigma_{ij}^{20} \Sigma_{ij}^{22} \\+ \alpha^{AB}_{20} \alpha^{AB}_{22}  \Sigma_{ij}^{22} \Sigma_{ij}^{20}.
\end{multline}
Inserting (\ref{3.229}) into the expression of strain energy density (\ref{b2}) and doing the volume integration gives
\begin{multline}
\label{3.228}
    E_{\rm{strain}} =  \mu \int  ((\alpha^{AB}_{20})^2 (\Sigma_{ij}^{20})^2 + (\alpha^{AB}_{22})^2 (\Sigma_{ij}^{22})^2 + \alpha^{AB}_{20} \alpha^{AB}_{22}  \Sigma_{ij}^{20} \Sigma_{ij}^{22} \\+ \alpha^{AB}_{20} \alpha^{AB}_{22}  \Sigma_{ij}^{22} \Sigma_{ij}^{20}) dV.
\end{multline}
The expressions of $ \Sigma_{ij}^{22}$ and $\Sigma_{ij}^{20}$ are given in appendix \ref{App1}. We get
\begin{equation}
\label{3.230}
    \Sigma_{ij}^{20} \Sigma_{ij}^{22}  =  \Sigma_{ij}^{22} \Sigma_{ij}^{20} = -\frac{3}{2}\sqrt{\frac{15}{2 \pi R^2}}\alpha_{20}^{AB} \alpha_{22}^{AB} r^2 \cos{2\phi}.
\end{equation}
This gives
\begin{equation}
    \int_V \Sigma_{ij}^{20} \Sigma_{ij}^{22} dV =  \int_V \Sigma_{ij}^{22} \Sigma_{ij}^{20} dV = 
 0.
\end{equation}
and (\ref{3.228}) becomes
\begin{equation}
\label{3.228a}
    E_{\rm{strain}} = E_{\rm{strain}}^{20} + E_{\rm{strain}}^{22} = 38 \pi (R^3 - R_c^3)\mu (\alpha^{AB}_{20})^2 + 5 (R^3 - R_c^3)\mu (\alpha^{AB}_{22})^2 .
\end{equation}
We therefore find that there are no cross strain energy terms when both perturbations are present. Therefore, the last term of (\ref{3.228}), which corresponds to the cross term in the strain energy $B_{20,22}(\epsilon^{SB}_{20} -\epsilon^{SA}_{20,0})(\epsilon^{SB}_{22} -\epsilon^{SA}_{22,0})$, is zero. The value of $E_{\rm{strain}}^{20}$ obtained here, when inserted into the equation (\ref{7}), gives the expression for $B_{20}$ (\ref{8}) which agrees with the one given in literature (\cite{Baym_Pines_1971}).

Next, we calculate the cross term for the perturbed gravitational potential energy corresponding to $A_{20,22} \epsilon^{SB}_{20}\epsilon^{SB}_{22}$ in the expression of the total energy (\ref{117}). For this we use the expression for perturbed gravitational energy (\ref{98}) from \cite{FRIEDMAN}. Using (\ref{100}) and (\ref{101}), we get the following expressions of the perturbed gravitational potential $\delta\Phi_{\rm{int}} (0 \leq r \leq R)$  and $\delta\Phi_{\rm{ext}} ( r > R )$ respectively,

\begin{equation}
\label{100c}
    \delta\Phi_{\rm{int}} = -\frac{8 \pi \rho G}{5} \alpha_{20} r^2 Y_{20}(\theta)  -\frac{8 \pi \rho G}{5} \alpha_{22} r^2 Y_{22}(\theta, \phi),
\end{equation}

\begin{equation}
\label{101c}
    \delta\Phi_{\rm{ext}} = -\frac{8 \pi \rho G}{5} \alpha_{20} \frac{R^5}{r^3} Y_{20}(\theta) -\frac{8 \pi \rho G}{5} \alpha_{22} \frac{R^5}{r^3} Y_{22}(\theta, \phi).
\end{equation}

Inserting the expression of displacement field of equation (\ref{118}), the unperturbed gravitational potential energy of equation (\ref{99}), and the perturbed gravitational potentials of equations (\ref{100c}) and (\ref{101c}) into (\ref{98}) then gives the perturbation in the gravitational potential energy.  The expression is large , so we will not write it down, but one easily finds that that there are no cross terms due to the orthogonality property of the spherical harmonics $Y_{20}$ and $Y_{22}$ when integrated over a 2-sphere.

\section{Ellipticity at equilibrium}
\label{section 6}

Given that there are no cross-terms in the expression for the star's energy, equation (\ref{117}) reduces to the sum of the separate spherical, $m=0$ and $m=2$ contributions:
\begin{multline}
\label{R3}
    E_{\rm{T}} = E_{\rm{grav},s} + \frac{L^2}{2I_{zz}} + B_{20}(\epsilon^{SB}_{20} - \epsilon^{SA}_{20,0})^2  + A_{20}(\epsilon^{SB}_{20})^2 \\ +B_{22}(\epsilon^{SB}_{22} - \epsilon^{SA}_{22,0})^2  + A_{22}(\epsilon^{SB}_{22})^2,
\end{multline}
Perturbation $l=2, m=2$ do not contribute to the kinetic energy because we have $\delta I^{22}_{zz} = 0$. One can find this result in the Section 8.5.1 of \cite{soton467267}.

It follows  that the $m=0$ and $m=2$ deformations are those obtained by analysing each case separately.  Explicitly, if one minimises $E_{\rm T}$ of equation (\ref{R3}) with respect to $\epsilon_{20}$, keeping $L$, $\epsilon_{20,0}$, $\epsilon_{22}$ and $\epsilon_{22,0}$ fixed, one obtains equation (\ref{11}) for $\epsilon_{20}$, containing the expected centrifugal and elastic deformation terms.  If instead one minimises $E_{\rm T}$  with respect to $\epsilon_{22}$, keeping $L$, $\epsilon_{20}$, $\epsilon_{20,0}$ and $\epsilon_{22,0}$ fixed, one obtains equation (\ref{11c}) for $\epsilon_{22}$,

Given equations (\ref{b20_b22}) and (\ref{a20_a22}), we can see that just as $B_{20} \ll A_{20}$, we also have $B_{22} \ll A_{22}$.  Using these relations we obtain
\begin{equation}
\label{eq:b_ratio}
    \frac{B_{22}}{A_{22}} = \frac{25}{19} \frac{B_{20}}{A_{20}} . 
\end{equation}
We point this out as it is these ratios that determine how large an actual deformation ($\epsilon_{20}$ or $\epsilon_{22}$) one gets for a given zero strain reference shape ($\epsilon_{20,0}$ or $\epsilon_{22,0}$).  

For both axisymmetric and non-axisymmetric deformations, the effects of elasticity are weak compared with gravity, and the actual deformation resulting from a given reference shape are small.  The numerical factor that appears in equation (\ref{eq:b_ratio}) is specific to our simple uniform density uniform shear modulus stellar model, but we expect the result $B_{22} / A_{22} \sim B_{20} / A_{20}$ will remain true for more realistic stellar models, providing the effects of elasticity remain small.

Inserting some numerical values for the case of mountains, we have
\begin{equation}
\label{R1}
    A_{22} = \frac{1}{25 R}G M^2 =  2.091 \times 10^{52} \rm{erg} \frac{M_{1.4}^2}{R_6},
\end{equation}
\begin{equation}
\label{R2}
    B_{22} =  \frac{1}{2} \mu V_{\rm{c}} .
\end{equation}
We can write $B_{22}$ in terms of $\Delta R$ as follows,
\begin{equation}
\label{R2_deltar}
    B_{22} =   2 \pi \mu  R^2 \Delta R  = 0.628 \times 10^{48} \rm{erg} \hspace{0.1cm} \mu_{30}  R_6^2 \Delta R_5  .
\end{equation}
For the mountain size itself, we have
\begin{equation}
\label{104}
    \epsilon^{SB}_{22} =  b_{22}\epsilon^{SA}_{22,0},
\end{equation}
where
\begin{equation}
\label{b22_num}
    b_{22} = \frac{B_{22}}{A_{22}+B_{22}} \approx \frac{B_{22}}{A_{22}} = \frac{50 \pi  \mu R^3   \Delta R}{M^2 G} \approx 3.005 \times 10^{-5} \frac{\mu_{30} R_{6}^3 \Delta R_5 }{M_{1.4}^2}  
\end{equation}
Substituting (\ref{b22_num})  into (\ref{104}) we get,
\begin{equation}
\label{105}
    \epsilon^{SB}_{22} \approx 3.005 \times 10^{-5} \epsilon^{SA}_{22,0} \frac{\mu_{30} R_{6}^3 \Delta R_5 }{M_{1.4}^2}.
\end{equation}
Expression (\ref{105}) quantifies the  smallness of the equilibrium ellipticity of the star relative to its  relaxed (zero-strain) star ellipticity.


\section{Change in total energy during starquake}
\label{energy_cal}

\begin{figure}
    \centering
    \includegraphics[width=0.35\textwidth]{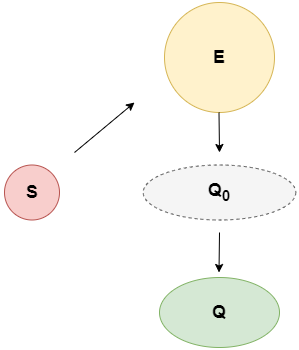}
    \caption{Schematic picture of the starquake model of Section \ref{energy_cal}. We are looking at the stars in the $x$-$y$ plane, with rotation along $Oz$. Star S (red) is the non-rotating elastic spherical star. It spins up and goes to the configuration E (yellow). At this point, the crust breaks and acquires a new equilibrium configuration Q (green). Q$_0$ (grey) is the relaxed shape for the new equilibrium shape Q.}
    \label{fig9}
\end{figure}

We will now apply our starquake model to a specific scenario, of the same sort considered by \citet{Fattoyev_2018} and \cite{13}, and described schematically in Fig. \ref{fig9}.

We start with a non-rotating spherical elastically relaxed star S. We spin it up to pre-quake equilibrium configuration E, modelling the rotation as a perturbation, as described in Section \ref{section 5}. The crust of star  E then fractures,  and the star acquires a new equilibrium shape Q, with star Q$_0$ being the relaxed configuration for the new equilibrium shape Q.

To build intuition, in Section \ref{cal_del_et_m_0} we consider fractures described by pure $m=0$ perturbations, while in Section \ref{cal_del_et_m_2} we consider fractures described by pure $m=2$ perturbations.   Then, in Section \ref{sect:both}, we consider the case in which we are really interested, with a combination of $m=0$ and $m=2$ perturbations.

Crucially, we enforce both angular momentum and energy conservation.  Angular momentum conservation is enforced by requiring that the angular momentum of star Q is the same as that of star E.  Energy conservation is enforced by requiring that the energy of star Q is equal to or less than that of star E, as a realistic fracture will generate heat within the star, and also emit gravitational waves, things which we do not explicitly include in our energy calculations.  Note that as long as the quake is axisymmetric, the gravitational wave emission will not carry any angular momentum (see e.g. \citet{Yim_2022}).

We take the view that star E is given, leaving us with a two-parameter family of possible quakes, corresponding to the $m=0$ and $m=2$ perturbations.  We take as our two free parameters the changes in relaxed shapes $\Delta \epsilon_{20,0}$ and $\Delta \epsilon_{22,0}$.  We wish to find the region in this parameter space for which starquakes are allowed, and how large a mountain can be formed.

\subsection{Perturbation $l=2, m=0$}
\label{cal_del_et_m_0}

The expression for the total energy of the star E, for pure $m=0$ perturbations, is given as (\ref{10}). To find the shape of star E, we minimise this energy as described previously, keeping the angular momentum $L$ and reference shape $\epsilon_{20,0}$ fixed.  This gives the usual relation (\ref{11}) between the equilibrium shape  and the elastically relaxed shape. However, in this case we have $\epsilon_{20,0}=0$, as star E was obtained by spinning up the spherical and relaxed star S, so (\ref{11}) reduces to
\begin{equation}
\label{ep}
     \epsilon_{20} = \frac{I_{zz,s} \Omega^2 }{4(A_{20}+B_{20})}.
\end{equation}

When the crust breaks, the axisymmetric perturbation ($\epsilon_{20}$) changes, on top of the star E. The moment of inertia and the angular velocity of the star E transforms as $I \rightarrow I+\Delta I$ and $\Omega \rightarrow \Omega +\Delta \Omega $ respectively. Since, the angular momentum is conserved, we have
\begin{equation}
\label{ang_mom_con}
    L = (I+\Delta I)(\Omega +\Delta \Omega) = I \Omega.
\end{equation}
Rearranging equation (\ref{ang_mom_con}) gives,
\begin{equation}
\label{frac_change}
    \frac{\Delta I}{I} = - \frac{\Delta \Omega}{\Omega} - \frac{\Delta I}{I}\frac{\Delta \Omega}{\Omega}.
\end{equation}
The change in the moment of inertia is given as
\begin{equation}
\label{del_i}
    \Delta I =  \Delta (I_{zz,s}(1+\epsilon_{20})) = I_{zz,s} \Delta \epsilon_{20}.
\end{equation}
Using (\ref{frac_change}) and (\ref{del_i}) we get,
\begin{equation}
\label{del_om}
    \frac{\Delta \Omega}{\Omega} = - \frac{\Delta \epsilon_{20}}{1+\epsilon_{20} +\Delta \epsilon_{20}}.
\end{equation}

To find the relation between the actual change in shape $\Delta \epsilon_{20}$ in going from star E to star Q, to the change in reference shape $\Delta \epsilon_{20,0}$, we perturb (\ref{11}). This gives
\begin{equation}
\label{del_ep}
    \Delta \epsilon_{20} = \frac{2 I_{zz,s} \Omega \Delta \Omega }{4(A_{20}+B_{20})} + \frac{B_{20}}{A_{20}+B_{20}} \Delta \epsilon_{20,0}.
\end{equation}
Using (\ref{ep}) and  (\ref{del_om}), (\ref{del_ep}) becomes
\begin{equation}
\label{del_ep_n}
    \Delta \epsilon_{20} = -\frac{2 \epsilon_{20} \Delta \epsilon_{20}}{1+\epsilon_{20} +\Delta \epsilon_{20}} + \frac{B_{20}}{A_{20}+B_{20}} \Delta \epsilon_{20,0}.
\end{equation}
The above equation is a quadratic in $\Delta \epsilon_{20}$. Using the standard formula for finding the roots of a quadratic and then performing linearization, we get the following relation between $\Delta \epsilon_{20}$ and $\Delta \epsilon_{20,0}$ to the leading order in b$_{20}$,
\begin{equation}
\label{del_ep_nn}
    \Delta \epsilon_{20} = b_{20} \Delta \epsilon_{20,0}(1- 2 \epsilon_{20}).
\end{equation}

\vspace{0.4cm}
The change in the kinetic energy between the two equilibrium configurations E and Q is given as
\begin{equation}
\label{del_ek}
    \Delta E_{\rm{k}} = \Delta\left(\frac{L^2}{2I_{zz,s}(1+\epsilon_{20})}\right)= - \frac{(I_{zz} \Omega)^2}{2I_{zz,s}(1+\epsilon_{20})^2}\Delta \epsilon_{20}.
\end{equation}
Inserting the expression of moment of inertia (\ref{n18}) and $\epsilon_{20}$ (\ref{ep}) into (\ref{del_ek}) we get,
\begin{equation}
\label{del_ek_1}
    \Delta E_{\rm{k}} = - \frac{I_{zz,s} \Omega^2}{2}\Delta \epsilon_{20} = -2 (A_{20}+B_{20}) \epsilon_{20}\Delta \epsilon_{20}.
\end{equation}

The change in the gravitational potential energy between the two equilibrium configurations is given as
\begin{equation}
\label{del_eg}
    \Delta E_{\rm{g}} = \Delta(A_{20} \epsilon^2_{20})=2A_{20} \epsilon_{20}\Delta \epsilon_{20}.
\end{equation}
We see that,
\begin{equation}
\label{del_eg_ek}
     \Delta E_{\rm{g}} +  \Delta E_{\rm{k}} = -2 B_{20} \epsilon_{20}\Delta \epsilon_{20}.
\end{equation}

To calculate the change in the strain energy, we cannot use the perturbation method, as used so far, because we will allow for order unity changes in strain energy, i.e.\ we will allow the star to loose all or most of its strain energy, corresponding to a single large fracture event. To calculate the change in the strain energy between the two equilibrium configurations, we will therefore simply calculate the strain energy of the respective configurations (E and Q) and take the difference:
\begin{equation}
\label{es_diff}
    \Delta E_{{\rm{s}}} = E^Q_{\rm{s}} - E^E_{\rm{s}} =  B_{20}(\epsilon^Q_{20} - \epsilon^Q_{20,0})^2 - B_{20}(\epsilon^E_{20} - \epsilon^E_{20,0})^2. 
\end{equation}
We have, 
\begin{equation}
\label{ep_q}
    \epsilon^Q_{20} = \epsilon_{20} + \Delta \epsilon_{20}
\end{equation}
and 
\begin{equation}
\label{ep_q_0}
  \epsilon^Q_{20,0} =  \epsilon_{20,0} + \Delta \epsilon_{20,0} = \Delta \epsilon_{20,0}.  
\end{equation}
Inserting (\ref{ep_q}) and (\ref{ep_q_0}) into (\ref{es_diff}), we get
\begin{equation}
\label{del_es}
    \Delta E_{\rm{s}} = B_{20}(\epsilon_{20} + \Delta \epsilon_{20} - \Delta \epsilon_{20,0})^2 - B_{20}\epsilon^2_{20}. 
\end{equation}
We insert (\ref{del_ep_nn}) into (\ref{del_es}) and further simplify it. This gives,
\begin{multline}
\label{del_es_1}
    \Delta E_{\rm{s}} = B_{20} \Delta \epsilon^2_{20,0} - 2 B_{20 } \epsilon_{20}  \Delta \epsilon_{20,0} + 2 B_{20} \epsilon_{20}\Delta \epsilon_{20}\\ + B_{20} \Delta \epsilon^2_{20} - 2B_{20} \Delta \epsilon_{20}\Delta \epsilon_{20,0}. 
\end{multline}

The change in the total energy for the $m=0$ perturbation is given as,
\begin{multline}
\label{del_et_0}
    \Delta E_{\rm{T}} = \Delta E_{\rm{k}} + \Delta E_{\rm{g}} + \Delta E_{\rm{s}} =   B_{20} \Delta \epsilon^2_{20,0} - 2 B_{20 } \epsilon_{20}  \Delta \epsilon_{20,0} \\ + B_{20} \Delta \epsilon^2_{20} - 2B_{20} \Delta \epsilon_{20}\Delta \epsilon_{20,0}.
\end{multline}
The change in the energy, corresponding to $\Delta E_{\rm{k}} + \Delta E_{\rm{g}}$ (\ref{del_eg_ek}), gets completely cancelled by the third term in the expression of $\Delta E_{\rm{s}}$ (\ref{del_es_1}). Also, the energy given by equation (\ref{del_eg_ek}) is $~b_{20}$ times smaller than the leading order terms in $\Delta E_{\rm{s}}$. Since, the third and fourth term in (\ref{del_et_0}) are order of magnitude smaller than first two terms, we can ignore them. This gives,

\begin{equation}
\label{del_et}
    \Delta E_{\rm{T}} = B_{20} \Delta \epsilon^2_{20,0} - 2 B_{20 } \epsilon_{20}  \Delta \epsilon_{20,0}.
\end{equation}
Therefore, to the leading order, the change in the total energy $ \Delta E_{\rm{T}} $ includes contribution only from $\Delta E_{\rm{s}}$. In (\ref{del_et}), the free parameter is $\Delta \epsilon_{20,0}$. The total change in energy will be negative as long as,
\begin{equation}
\label{eq:maximum_Delta_epsilon_20}
   \Delta \epsilon_{20,0} < 2 \epsilon_{20} . 
\end{equation}
This has a simple physical interpretation.  For very small values of $\Delta \epsilon_{20,0}$ the strain energy is guaranteed to decrease, as some of the strain created by spinning up the spherical star is relieved.  All of this strain would be relieved if $\Delta \epsilon_{20,0} = \epsilon_{20}$, as the new reference shape would match the actual shape of the star.  If the star ``overshoots'', less energy is relieved.  In the case of overshooting as far as $\Delta \epsilon_{20,0} \leq 2 \epsilon_{20}$, the new strain is equal in magnitude to the pre-quake strain, but is acting to make the star more, not less, oblate, and the change is energetically neutral, in terms of strain energy.  

\begin{figure}
    \centering
    \includegraphics[width=0.45\textwidth]{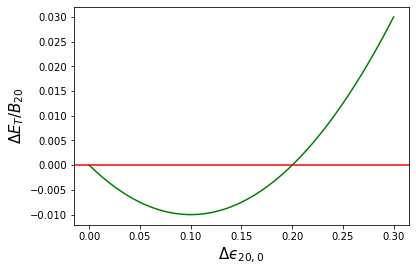}
    \caption{Change in energy for an axisymmetric star that undergoes a purely axisymmetric change in reference shape of size $\Delta \epsilon_{20,0}$, as described in Section \ref{cal_del_et_m_0}. We have set $\epsilon_{20} = 0.1 $ in this plot. }\label{fig10}   
\end{figure}

For a given $\epsilon_{20}$, the variation of $\frac{\Delta E_{\rm{T}}}{B_{20}} $ with $\Delta \epsilon_{20,0}$ is shown in Fig. \ref{fig10}. We have set $\epsilon_{20} = 0.1$ corresponding to the fastest rotating pulsar (716 revolutions/s). One can chose any value of $\epsilon_{20}$ and the curve in Fig. \ref{fig10} will scale accordingly.

Also, we verified that if we include 3rd order terms in the expression of the total energy as shown below,
\begin{multline}
\label{total_e_order3}
   E_{\rm{T}} = E_{\rm{grav},s} + \frac{L^2}{2I_{zz,s}(1+\epsilon_{20})} +   A_{20}\epsilon_{20}^2
    + B_{20}(\epsilon_{20} - \epsilon_{20,0})^2 \\ + A'_{20}\epsilon_{20}^3
    + B'_{20}(\epsilon_{20} - \epsilon_{20,0})^3,
\end{multline}
the correction terms in $ \Delta E_{\rm{T}}$ would be of higher order. Therefore, we can safely assume (\ref{10}) as our model of analysis.

\subsection{Perturbation $l=2, m=2$}    \label{cal_del_et_m_2}

We will now calculate the change in the total energy for a pure $m=2$ perturbation of star E. The expression of the total energy of the pre-quake equilibrium star E is given by equation (\ref{10}). When the starquake occurs, star E acquires a new equilibrium configuration Q. At the starquake, we add only the non-axisymmetric perturbation ($\epsilon_{22}$) on top of the star E. The expression of the total energy of the new equilibrium star Q is given as,
\begin{multline}
\label{total_e_2}
   E_{\rm{T}} = E_{\rm{grav},s} + \frac{L^2}{2I_{zz,s}(1+\epsilon_{20})} + A_{20}\epsilon_{20}^2
    + B_{20}(\epsilon_{20} - \epsilon_{20,0})^2 \\ +   A_{22}\epsilon_{22}^2
    + B_{22}(\epsilon_{22} - \epsilon_{22,0})^2.
\end{multline}
Here, $\epsilon_{22,0}$ is the relaxed ellipticity of star Q$_0$. To get the relation between the new equilibrium Q and its relaxed shape Q$_0$ of the star for this $m=2$ perturbation, we minimise (\ref{total_e_2}) w.r.t. $\epsilon_{22}$ at fixed $L$, $\epsilon_{20}$, $\epsilon_{20,0}$ and $\epsilon_{22,0}$.  This gives
\begin{equation}
\label{ep_2}
    \epsilon_{22} =  \frac{B_{22}}{A_{22}+B_{22}}\epsilon_{22,0} \approx \frac{B_{22}}{A_{22}}\epsilon_{22,0}.
\end{equation}
Perturbing (\ref{ep_2}) to describe the change in going from E to Q gives
\begin{equation}
\label{del_ep_2}
    \Delta \epsilon_{22} =  \frac{ B_{22}}{ A_{22} + B_{22} } \Delta \epsilon_{22,0} = b_{22} \Delta \epsilon_{22,0},
\end{equation}

The change in the strain energy stored in the crust between the two equilibrium configurations E and Q, for pure $m=2$ perturbation is given as
\begin{equation}
\label{es_diff_22}
    \Delta E_{\rm{s}} =  B_{22}(\epsilon_{22} - \epsilon_{22,0})^2.
\end{equation}
Since, there is no mountain before the starquake, we can write the equilibrium shape ($\epsilon_{22}$) and the relaxed shape ellipticity ($\epsilon_{22,0}$) of star Q as the change in the equilibrium ($\Delta \epsilon_{22}$)  and relaxed shape ellipticity ($\Delta \epsilon_{22,0}$) between star E and Q, i.e.,
\begin{equation}
\label{ep_q_22}
   \Delta \epsilon_{22} =  \epsilon^Q_{22} - \epsilon^E_{22} = \epsilon_{22} - 0 = \epsilon_{22}
\end{equation}
and 
\begin{equation}
\label{ep_q_0_22}
  \Delta \epsilon_{22,0} =  \epsilon^Q_{22,0} - \epsilon^E_{22,0} = \epsilon_{22,0} - 0 = \epsilon_{22,0}
\end{equation}
Inserting (\ref{ep_q_22}) and (\ref{ep_q_0_22}) into (\ref{es_diff_22}), we get
\begin{equation}
    \Delta E_{\rm{s}} =   B_{22}(\Delta \epsilon_{22}- \Delta \epsilon_{22,0})^2 \approx B_{22} \Delta  \epsilon^2_{22,0} .
\end{equation}

The change in the gravitational potential energy, between the two equilibrium configurations E and Q, for pure $m=2$ perturbation is given as,
\begin{equation}
    \Delta E_{\rm{g}} = A_{22} \epsilon^2_{22} =  A_{22} \Delta \epsilon^2_{22}.
\end{equation}
Using (\ref{del_ep_2}) , we get
\begin{equation}
    \Delta E_{\rm{g}} = B_{22} b_{22} \Delta \epsilon^2_{22,0}.
\end{equation}
The change in the gravitational potential energy  $ \Delta E_{\rm{g}}$ is $~b_{22}$ order of magnitude smaller than $ \Delta E_{\rm{s}}$. Therefore, the change in the total energy is given as,
\begin{equation}
\label{del_et_22_final}
    \Delta E_{\rm{T}} \approx \Delta E_{\rm{s}} \approx B_{22} \Delta  \epsilon^2_{22,0}  . 
\end{equation}

Since the change in the total energy of the star is  positive, we see that it is not possible for a star to have a pure $m=2$ perturbation. It follows that any $m=2$ perturbation must always be present with a axisymmetric $m=0$ perturbation.  We go on to consider this case in Section \ref{sect:both}.

\subsection{Perturbation $m=0$ and $m=2$ present together} \label{sect:both}

Previously in Section \ref{section 6} we showed that when both the perturbations $m=0$ and $m=2$ are present together, the changes in the energies of the star for the respective perturbations are independent of each other. Therefore, to obtain the expression of the change in the total energy, when both the perturbations are present, we can simply add the results (\ref{del_et}) and (\ref{del_et_22_final}) given in Section \ref{cal_del_et_m_0} and Section \ref{cal_del_et_m_2} respectively. This gives
\begin{equation}
\label{del_et_0_2}
    \Delta E_{\rm{T}} = B_{20} \Delta \epsilon^2_{20,0} - 2 B_{20 } \epsilon_{20}  \Delta \epsilon_{20,0} + B_{22} \Delta  \epsilon^2_{22,0} 
\end{equation}

\begin{figure}
    \centering
    \includegraphics[width=0.45\textwidth]{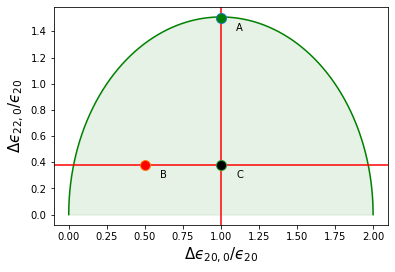}
    \caption{Illustration of the region in our $(\Delta \epsilon_{20,0}, \Delta \epsilon_{22,0})$ parameter space for which $\Delta E_{\rm{T}} < 0$. The green curve marks the elliptical boundary of this region, where $\Delta E_{\rm T}=0$; energetically allowed points lie on and below this curve, in the green shaded region. We give the position of three different configurations of the star. Point A (green) marks the position of the star with largest mountain. The vertical red line gives the configurations of the star which releases maximum amount of energy while forming a mountain. This is shown explicitly in one of the cases as shown by the horizontal red line where point B (red) indicates the position of the star with a mountain size as discussed in a recent study \citet{13} and point C (black) gives the position of the star configuration with the same mountain as star B, but which requires the least energy.}
    \label{fig12}
\end{figure}

Here, we have two free parameters, $\Delta \epsilon_{20,0}$ and  $\Delta \epsilon_{22,0}$. Fig. \ref{fig12} shows (in green)  the curve in the 2-d parameter space of $\Delta \epsilon_{20,0}$ and  $\Delta \epsilon_{22,0}$ when $\Delta E_{\rm{T}} =0$. For a mountain to form, we need
\begin{equation}
\label{del_et_0_2_con}
    \Delta E_{\rm{T}} = B_{20} \Delta \epsilon^2_{20,0} - 2 B_{20 } \epsilon_{20}  \Delta \epsilon_{20,0} + B_{22} \Delta  \epsilon^2_{22,0} < 0,
\end{equation}
which is the region inside the green curve. We can re-write (\ref{del_et_0_2}) in the form given as,
\begin{equation}
\label{et_curve}
    \frac{(\Delta \epsilon_{20,0} - \epsilon_{20})^2}{\epsilon_{20}^2} + \frac{B_{22} (\Delta \epsilon_{22,0})^2}{\epsilon_{20}^2 B_{20}} = 1.
\end{equation}
This clearly shows that the curve (green) is an ellipse.  Note that we can, without loss of generality, consider only the upper half of the ellipse, as we can always map $\epsilon_{22} < 0$ to $\epsilon_{22} > 0$ through a $\pi/2$ rotation in the $x$-$y$ plane; see e.g. \citet{jones_2015}. 

The form of this curve can be easily understood.  As demonstrated in Section \ref{cal_del_et_m_2}, when $\Delta \epsilon_{20,0}$ is zero, it isn't energetically possible to form a mountain, so $\Delta \epsilon_{22,0}=0$; some of the strain energy stored in the axisymmetric deformation has to be liberated to build the mountain.  This explains why the curve starts at the origin.  As $\epsilon_{20,0}$ then increases, the liberated axisymmetric strain energy can then be used to build a mountain, so $\Delta \epsilon_{22,0}$ increases also.  The liberated axisymmetric strain energy is a maximum when $\Delta \epsilon_{20,0} = \epsilon_{20}$, as described in Section \ref{cal_del_et_m_2} (see discussion following equation (\ref{eq:maximum_Delta_epsilon_20})), so this is where $\epsilon_{22,0}$ peaks.  As $\Delta \epsilon_{20,0}$ is increased further, the axisymmetric strain energy liberated in the quake decreases, decreasing the mountain $\Delta \epsilon_{22,0}$, until we reach $\Delta \epsilon_{20,0} = 2 \epsilon_{20}$, where the liberated axisymmetric strain energy is zero, and no mountain can be built (as per equation (\ref{eq:maximum_Delta_epsilon_20})).


\section{Analysis and interpretation}
\label{2-d space}

We will now examine some of the consequences of our model.  In Section \ref{sect:max_mountain} we talk about maximum mountain sizes, while in Section \ref{sect:GC} we show how the recent analysis of \citet{13} can be described within our framework. 

Throughout this Section, we will continue to imagine we start with a non-rotating elastically relaxed spherical star S that is then spun up to some angular velocity $\Omega$ to give the
rotating, elastically strained star E, with a given oblateness $\epsilon_{20}$, which then undergoes a starquake to form a triaxial star Q.  We will  continue to use $\Delta \epsilon_{20,0}$ and $\Delta \epsilon_{22,0}$ as the two free parameters describing the quake, i.e.\ the $m=0$ and $m=2$ changes in reference oblateness and ellipticity.

\subsection{The maximum mountain} \label{sect:max_mountain}

Before calculating the largest mountain than can form when star E undergoes a quake, we calculate the most energetically favoured change in axisymmetric configuration when creating a mountain \emph{of given size} $\epsilon_{22}$.  This is obtained by minimising the energy perturbation $\Delta E_{\rm T}$ of equation (\ref{del_et_0_2}) at fixed $\Delta \epsilon_{22,0}$.  This gives,
\begin{equation}
\label{eq:red_line}
    \Delta \epsilon_{20,0} = \epsilon_{20}.
\end{equation}
This simply says that the most energy is liberated if the $m=0$ reference shape increases by an amount $\Delta \epsilon_{20,0}$ equal to the original oblateness $\epsilon_{20}$ of star E.  This is shown as the vertical red line $\Delta \epsilon_{20,0} = \epsilon_{20}$  in Fig. \ref{fig12}.

The largest mountain that can be formed in the starquake corresponds to the topmost point of the ellipse, shown as the green point and labelled as star A in Fig. \ref{fig12}.  Setting $\Delta \epsilon_{20,0} = \epsilon_{20}$ and $\Delta E_{\rm T} = 0$ in equation (\ref{del_et_0_2}), we obtain
\begin{equation}
    \Delta \epsilon_{22,0}^{\rm max} = \sqrt{\frac{B_{20}}{B_{22}}}\epsilon_{20}.
\end{equation}
This is the reference (i.e. zero strain) shape for this mountain.  To obtain the actual mountain size we can use equation (\ref{del_ep_2}), obtaining
\begin{equation}
\label{max_moun}
    \Delta \epsilon_{22}^{\rm max} = \frac{\sqrt{B_{22}B_{20}}}{A_{22}} \epsilon_{20}.
\end{equation}
This is our main result; it is the largest mountain that one can build when star E undergoes a starquake, subject to the constraints of angular momentum conservation and energy conservation.  Note the presence of the 
geometric mean of $B_{20}$ and $B_{22}$ in the expression of $\Delta \epsilon_{22}^{\rm max}$.  The building of an $m=2$ mountain necessarily requires a release of $m=0$ strain, coupling the two sorts of deformation together.

\subsection{Description of \citet{13}} \label{sect:GC}

As a further application of our model, we now use it to describe a particular mountain-building starquake scenario presented in \cite{13}.  Our formalism will allow us to explore a few things not considered in \citet{13}.  We will show that the model of \citet{13} involves not only an $m=2$ change, but necessarily also an $m=0$ change.  We will be able to calculate the corresponding reference shape the post-quake star would have, i.e.\ how its zero-strain configuration is reconfigured in the quake, for both the $m=0$ and $m=2$ perturbations.  We will also confirm that (at fixed angular momentum) the energy of the star does indeed decrease, and how close the \citet{13} mountain is to the maximum possible mountain that the quake process allows, as per equation (\ref{max_moun}) above.

\begin{figure}
    \centering
    \includegraphics[width=0.45\textwidth]{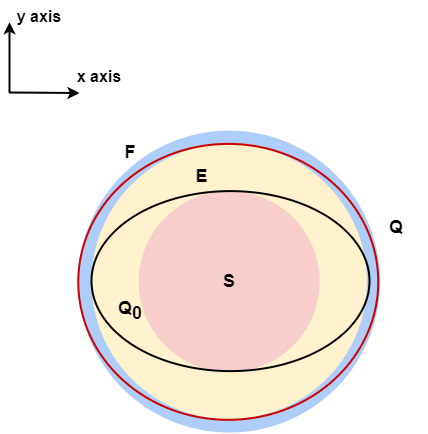}
    \caption{Schematic depiction of various configurations of the spinning up star in $x$-$y$ plane, with rotation along the $z$-axis. The pink disk S represents the spherical non-rotating zero strain configuration. The yellow disk E represents the elastic strained axisymmetric configuration just before the glitch. The blue disk F represents a star with the same angular momentum as E, but elastically relaxed and axisymmetric (i.e. equivalent to a fluid). The red ellipse Q represents the post-glitch equilibrium shape. The black ellipse Q$_0$ represents the elastically relaxed shape corresponding to star Q. We assume that the moment of inertia along the x-axis is smaller than the moment of inertia along the y-axis for both Q and Q$_0$ configurations.  The differences in shape of the various configurations are greatly exaggerated for clarity. }
    \label{figB}
\end{figure}

\citet{13} considered several different stellar configurations, which we will now describe. We display them schematically in Fig. \ref{figB}, where we show the cross-section in the $x$-$y$ plane, with, as always, rotation along the $z$-axis. We greatly exaggerate the size differences between different stars for clarity.  

The non-rotating elastically relaxed spherical star S is shown as a pink disk.  This is then spun-up to give the star that exists just prior to the starquake.  This is star E, and is rotating, elastically strained and axisymmetric, with moment of inertia tensor $\Delta I^E =$ Diag$(\Delta I_{xx}^E, \Delta I_{xx}^E,\Delta I_{zz}^E )$.  It is shown as the yellow disk.  If all of the strain in star E is relieved (at fixed angular momentum), through either a series of starquakes, or through some plastic creep, one would obtain a configuration corresponding to a fluid star F, with a slightly larger oblateness than star E.  This is shown as the blue disk, and is rotating, unstrained, and axisymmetric, with the moment of inertia tensor $\Delta I^F =$ Diag$(\Delta I_{xx}^F, \Delta I_{xx}^F,\Delta I_{zz}^F )$.

\citet{13} wrote the post-quake configuration as $\Delta I^Q =$ Diag$(\Delta I_{xx}^Q, \Delta I_{yy}^Q,\Delta I_{zz}^Q )$, which we show as the red ellipse Q in Fig. \ref{figB}.  Note that, simply for the sake of definiteness, we have assumed the symmetry-breaking is such that the radius along the $x$-axis is larger than the radius along the $y$-axis, so that $\Delta I^Q_{xx} < \Delta I^Q_{yy}$.  

\citet{13} argued that the post-quake configuration is expected to lie between the elastic and fluid configurations, as sketched in Fig. \ref{figB}, i.e.\ the red ellipse Q must lie between the pre-glitch elastically strained configuration E (yellow disk) and the zero-strain fluid configuration F (blue disk). The authors concluded that the maximum value of $\Delta I_{yy}^Q - \Delta I_{xx}^Q$ is $\Delta I_{xx}^E - \Delta I_{xx}^F$. This corresponds to the red ellipse of Fig. \ref{figB} just touching its bounding disks E and F.  This was computationally useful, as the stellar configurations E and F can both be computed as perturbations away from star S.  \citet{13} did this numerically for SLy and BSk21 equations of state, while we do so analytically, using the displacement vector field (\ref{d1}), valid for our uniform density incompressible stars.

The radii of star E are given by the equations presented in Section \ref{section 5}, by setting the starting angular velocity to zero (corresponding to the non-rotating relaxed star S) and the final angular velocity to $\Omega$, the angular velocity of star E.  Equation (\ref{a_1_elastic}) gives
\begin{equation}
    a_1^{\rm E} = a_2^{\rm E} = R \left[ 1 + \frac{1}{2} \alpha_{20} \right]
\end{equation}
where, from equation (\ref{d2})
\begin{equation}
\label{eq:alpha_20_E}
    \alpha_{20} = \frac{\lambda}{1+b_{20}} \Omega^2 .
\end{equation}
The radii of star F are of the same form, but with departures from star S a factor $(1+b_{20})$ larger (as per equation (\ref{d2})):
\begin{equation}
    a_1^{\rm F} = a_2^{\rm F} = R \left[ 1 + \frac{1}{2} \alpha_{20} (1+b_{20}) \right]
\end{equation}

In the model of \cite{13}, we can now immediately write down the radii of star Q.  Along the $x$-axis, it has the radius of star F, so that
\begin{equation}
a_1^{\rm Q} = a_1^{\rm F} = R \left[ 1 + \frac{1}{2} \alpha_{20} (1+b_{20}) \right] ,
\end{equation}
while along the $y$-axis, star Q has the same radius as star E:
\begin{equation}
a_2^{\rm Q} = a_2^{\rm E} = R \left[ 1 + \frac{1}{2} \alpha_{20} \right] ,
\end{equation}

Given the radii of the equilibrium shape Q, we can easily calculate the oblateness $\epsilon_{20}^Q$ and ellipticity $\epsilon_{22}^Q$.  Using equation (\ref{eq:epsilon_20_radii}) we have
\begin{equation}
    \epsilon_{20}^Q = \alpha_{20}+ \frac{b_{20}\alpha_{20}}{2} + \frac{\alpha^2_{20}}{4}
\end{equation}
while equation (\ref{eq:epsilon_22_radii}) gives
\begin{equation}
    \epsilon_{22}^Q \approx \frac{b_{20}\alpha_{20}}{2}.
\end{equation}

Of most interest here are the \emph{changes} in $\epsilon_{20}^Q$ and $\epsilon_{22}^Q$, in going from star E to star Q.  The oblateness ($m=0$) of star E is given by equation (\ref{ep_alpha}) (with $\alpha_{20}$
 given by equation (\ref{eq:alpha_20_E})), so we have:
\begin{equation}
\label{del_ep_20q}
    \Delta \epsilon_{20} = \epsilon_{20}^Q - \epsilon_{20}^E =
    \frac{1}{2} b_{20}\alpha_{20} \approx \frac{1}{2} b_{20} \epsilon_{20} .
\end{equation}

The ellipticity ($m=2$) of  star E is, by assumption, zero, so we have
\begin{equation}
\label{GC_moun}
    \Delta \epsilon_{22} = \epsilon_{22}^Q - \epsilon_{22}^E = \epsilon_{22}^Q = \frac{b_{20}\alpha_{20}}{2} \approx \frac{1}{2} b_{20} \epsilon_{20} .
\end{equation}
In both equations (\ref{del_ep_20q}) and (\ref{GC_moun}) we have approximated equation (\ref{alpha_ep}) as $\alpha_{20} \approx \epsilon_{20}$.

This is a simple result: the mountain formation scenario of \citet{13} consists of equal increases in $\epsilon_{22}$ (creating the mountain) and in $\epsilon_{20}$ (releasing some axisymmetric strain).  We record these results in the middle column of Table \ref{t2}.

Having calculated the changes in equilibrium shape, as parameterised by $\Delta \epsilon_{20}$ and $\Delta \epsilon_{22}$, we can easily calculate the changes in the corresponding reference shapes, $\Delta \epsilon_{20,0}$ and $\Delta \epsilon_{22,0}$.  For the axisymmetric change, we can use equation (\ref{del_ep_nn}) to give
\begin{equation}
    \Delta \epsilon_{20,0} = \frac{1}{2} (\epsilon_{20} - \frac{\epsilon^2_{20}}{4})
\end{equation}
while for the non-axisymmetric change we can use equation (\ref{del_ep_2}) to give
\begin{equation}
\label{delta_ep22}
    \Delta \epsilon_{22,0} = \frac{b_{20}}{2b_{22}} (\epsilon_{20} - \frac{\epsilon^2_{20}}{4}).
\end{equation}
Using (\ref{b20_b22}) and (\ref{a20_a22}) we have
\begin{equation}
\label{b20_22}
    \frac{b_{20}}{b_{22}} = \frac{B_{20} A_{22}}{A_{20} B_{22}}=  \frac{19}{25}.
\end{equation}
so that 
\begin{equation}
    \Delta \epsilon_{22,0} = \frac{19}{50} (\epsilon_{20} - \frac{\epsilon^2_{20}}{4}).
\end{equation}
We record these values in column B of Table \ref{t2}.  

We can now plot the position of star B in our 2-d parameter space; see the red point in Fig. \ref{fig12}.  Reassuringly, star B lies inside the green bounding ellipse, so does indeed correspond to a decrease in the star's energy.  We can in fact compute the change in energy, $\Delta E_{\rm T}$, simply by plugging our results for  $\Delta \epsilon_{20,0}$ and  $\Delta \epsilon_{22,0}$ into equation (\ref{del_et_0_2}).  The resulting (negative) expression is given in column B, bottom row, of Table \ref{t2}. 

We can compare the size of the mountain formed in the \citet{13} process with the size of the maximum mountain, constrained only by angular momentum and energy conservation.  Taking the ratio of the $ \epsilon_{22}^{\rm{max}}$ of equation (\ref{max_moun}) with the mountain size of equation (\ref{GC_moun}), denoted below as $\epsilon_{22}^{\rm G.C.}$:
\begin{equation}
    \frac{  \epsilon_{22}^{\rm{max}}}{  \epsilon_{22}^{\rm{G.C.}}} = 2 \frac{A_{20}}{A_{22}} \sqrt{\frac{B_{22}}{B_{20}}} \left( \frac{1}{1- \frac{\epsilon_{20}}{4}}\right) \approx 6 \sqrt{\frac{25}{57}} \left(1+ \frac{\epsilon_{20}}{4} \right)\approx 3.97.
\end{equation}
This shows, to the leading order, the maximum mountain size $\epsilon_{22}^{\rm{max}}$ is approximately 4 times larger than the mountain $\epsilon_{22}^{\rm{G.C.}}$ built in the \citet{13} scenario.

As a final application of our model, we can consider a star C, defined to have the same mountain $\epsilon_{22}$ as that of \cite{13} (equation (\ref{GC_moun})), but lying on the red vertical maximal energy-release line of Fig. \ref{fig12}. We record its parameters in the final column of Table \ref{t2}.  Such a star has the same $\Delta \epsilon_{22}$ and the same $\Delta \epsilon_{22,0}$ as star B.  However, it has $\Delta \epsilon_{20,0} = \epsilon_{20}$, as per equation (\ref{eq:red_line}).  Using equation (\ref{del_ep_nn}) this gives $\Delta \epsilon_{20} = b_{20} \epsilon_{20}$.  These values can then be substituted into equation (\ref{del_et_0_2}) to give the total energy change $E_{\rm T}$.  Note that, as expected, the energy release in forming star C is larger (in magnitude) than that released in forming star B.

\begin{table*}
\Large
  \begin{tabular}{|l|l|l|l|l|}
   \hline
    & A & B & C   \\
    \hline 
    $\Delta \epsilon_{20}$ &   $ b_{20}\epsilon_{20}$ &  $\frac{b_{20}}{2} (\epsilon_{20} - \frac{\epsilon^2_{20}}{4})$ & $b_{20}\epsilon_{20}$    \\
    \hline 
    $\Delta \epsilon_{22}$ &   $b_{22}\epsilon_{20} \sqrt{\frac{B_{20}}{B_{22}}}$ &  $\frac{b_{20}}{2} (\epsilon_{20} - \frac{\epsilon^2_{20}}{4})$& $\frac{b_{20}}{2} (\epsilon_{20} - \frac{\epsilon^2_{20}}{4})$   \\ 
    \hline
     \hline 
    $\Delta \epsilon_{20,0}$ &   $ \epsilon_{20}$ &  $\frac{1}{2} (\epsilon_{20} - \frac{\epsilon^2_{20}}{4})$ & $\epsilon_{20}$   \\
    \hline 
    $\Delta \epsilon_{22,0}$ &   $\epsilon_{20} \sqrt{\frac{B_{20}}{B_{22}}}$ &  $\frac{19}{50} (\epsilon_{20} - \frac{\epsilon^2_{20}}{4})$& $\frac{19}{50} (\epsilon_{20} - \frac{\epsilon^2_{20}}{4})$  \\
    \hline  
      $\Delta E_{\rm{T}}$  &   $0 $ &  $-\frac{1957 B_{22}}{1250} \epsilon_{20}^2$ & $ -\frac{5339 B_{22}}{2500} \epsilon_{20}^2$ 
      \\
    \hline 
  \end{tabular}  
  \caption{ Summary of changes in oblateness,  ellipticity, and energy for the star configurations A, B and C as  described in Section \ref{2-d space}, and illustrated in Figure \ref{fig12}. Rows 1 and 2  gives the change in the equilibrium oblateness  ($\Delta \epsilon_{20}$) and ellipticity ($\Delta \epsilon_{22}$) for the equilibrium shape Q for the three different configurations A, B, and C. Similarly,  rows 3 and 4 gives the change in the relaxed shape oblateness ($\Delta \epsilon_{20,0}$) and ellipticity ($\Delta \epsilon_{22,0}$) for the equilibrium shape Q$_0$. Lastly row 5 gives the change in the total energy during the starquake. \label{t2} }
\end{table*}

\section{Summary}
\label{sec:summary}

We have extended the \cite{Baym_Pines_1971} energy-based analysis of starquakes, 
originally developed for glitches in axisymmetric stars, to allow for non-axisymmetric shape changes.  We followed  \citet{Baym_Pines_1971} in using a very simple stellar model, with uniform density and uniform crustal shear modulus, which allowed for a fully analytic treatment.  We modelled the  non-axisymmetric shape change as a simple $(l=m=2)$ vector spherical harmonic, supplemented by a spherical $(l=m=0)$ piece in order to give volume conservation to second order in the size of the perturbation.  This allowed us to compute the changes in energy (summing over kinetic, gravitational and elastic strain energy contributions) between different stellar configurations.  We showed that there are no cross terms in the expression of the total energy of the star when both the symmetrical $(l=2, m=0)$ and asymmetrical $(l=2, m=2)$ perturbations are present simultaneously.

We verified that the effect of elasticity is small for non-axisymmetric crustal strains, just as it is for axisymmetric ones.  Quantitatively, the size of 
the non-axisymmetric ellipticity $\epsilon_{22}$ for a star whose zero strain shape is $\epsilon_{22,0}$ is given by 
\begin{equation}
    \epsilon_{22} \approx \frac{B_{22}}{A_{22}}\epsilon_{22,0} \approx 3.005 \times 10^{-5} \epsilon_{22,0} \frac{\mu_{30} R_{6}^3 \Delta R_5 }{M_{1.4}^2},
\end{equation}
i.e.\ the crust has to have a very large zero strain distortion to produce a significant gravitational wave emitting mountain.  This result was expected, but the values of the coefficients $A_{22}$ and $B_{22}$ specific to the non-axisymmetric case had not been computed previously, only their axisymmetric equivalents $A_{20}$ and $B_{20}$.

As an application of our formalism, we described the case of a non-rotating spherical elastically relaxed star being spun up and then undergoing a single large starquake, as considered in \citet{Fattoyev_2018} and \citet{13}.  The quake was described by two free parameters: $\Delta \epsilon_{20,0}$ and  $\Delta \epsilon_{22,0}$, the changes in zero strain shape of the $m=0$ and $m=2$ perturbations, respectively. We found the region in this parameter space consistent with angular momentum and energy conservation ($\Delta J = 0, \Delta E < 0$).  We showed that the largest mountain that can be built, subject only to these constraints, is given by:
\begin{equation}
\label{max_moun_2}
    \epsilon_{22}^{\rm max} = \frac{\sqrt{B_{22}B_{20}}}{A_{22}} \epsilon_{20} = \frac{\sqrt{57} }{5 } \frac{B_{22}}{A_{22}} \epsilon_{20},
\end{equation}
where $\epsilon_{20}$ is the rotational oblateness of the pre-quake star. We showed that this is approximately a factor of $4$ larger than the mountain built in the particular fracture scenario explored in \citet{13}. We also showed that the formation of a mountain always requires a change in axisymmetric shape too; some axisymmetric energy has to be given up in order to build the mountain. 

To get a rough estimate of the maximum mountain size, we can insert $\epsilon_{\Omega}$ of equation (\ref{ep_omega}) into equation (\ref{max_moun_2}), as in our model this oblateness comes from the centrifugal forces in the spinning star: 
\begin{equation}
   \label{max_moun_omega}
    \epsilon_{22}^{\rm max} = \frac{\sqrt{57} }{5 } \frac{B_{22}}{A_{22}} \epsilon_{\Omega} \approx 7.98 \times 10^{-8} \left  (\frac{f}{100 \hspace{0.1 cm} \rm{Hz}} \right)^2 \frac{\mu_{30} R_{6}^6 \Delta R_5 }{M_{1.4}^3} . 
\end{equation}
This result is to be interpreted as follows: if an initially non-rotating, elastically strained star is spun up to rotation rate $f$, and then undergoes  a crustquake, it is the maximum mountain that can be formed.  

There is however a limit to the applicability of this result, as the oblateness in the spinning up star, and the mountain formed in the quake event, will be limited by the crust's finite breaking strain.  The maximum mountain size imposed by the finite breaking strain has been examined many times previously (\cite{10.1111/j.1365-2966.2006.10998.x}; \cite{PhysRevD.88.044004}; \cite{Gittins_2020}; \cite{10.1093/mnras/stab2048}; \cite{10.1093/mnras/stac3058} ), but in the context of our simple model, we can set the ellipticity at the time of the crustquake to the breaking strain $\sigma_{\rm break}$.  Setting $\epsilon_\Omega = \sigma_{ \rm break}$ in equation (\ref{ep_omega}) then gives the maximum frequency $f_{\rm break}$ that the star can be spun up to before fracture:
\begin{equation}
    f_{\rm break} \approx 753{\,\rm Hz} \left (\frac{\sigma_{\rm break}}{10^{-1}}\frac{M_{1.4}}{R^3_6} \right)^{\frac{1}{2}} .
\end{equation}
We have parameterised in terms of a breaking strain of $0.1$, motivated by the high levels of braking strain found in the molecular dynamics calculations of \citet{hk_09}.  The corresponding upper limit on the mountain can be obtained by substituting this into the second equality of equation (\ref{max_moun_omega}), or, more directly, by setting $\epsilon_\Omega = \sigma_{\rm break}$ in the first equality:
\begin{equation}
    \epsilon_{22, \, \rm break}^{\rm max} = 1.13 \times 10^{-5} \left  (\frac{ \sigma_{ \rm break}}{10^{-1} } \right) \frac{\mu_{30} R_{6}^3 \Delta R_5 }{M_{1.4}^2}
\end{equation}
This is the largest mountain that could be produced if an initially non-rotating star is spun up all the way to the point where the strain in its crust reaches the breaking strain, and then a crustquake occurs forming the largest possible mountain.

Clearly, our model is highly idealised, and needs to be improved in several ways, as we now discuss. 

(i)  We assumed all the strain in the crust to be coming from the centrifugal force, but from realistic glitch models we know that the neutron superfluid component is important too. Superfluid vortices, when pinned, give rise to strain in the crust due to the Magnus force acting on the vortices. A more accurate model could be built by considering the strain in the crust to be coming from both centrifugal and Magnus forces \citep{1991ApJ...366..261R}. 

(ii)  We assumed a star with an incompressible fluid core with uniform density and uniform shear modulus of the outer crust.   This model should be improved by considering a realistic EOS. 

(iii) To model the non-axisymmetric perturbation we chose a "Kelvin mode" displacement field which is valid only for an incompressible fluid star; exploration of a range of non-axisymmetric quadrupolar deformation would be worthwhile, ideally motivated by a more detailed analysis of how the  crust actually fails. 

(iv)  We examined the mountain formation under the Newtonian framework. Further refinement could be achieved by examining it within the relativistic framework (see e.g.\ \cite{10.1093/mnras/stab2048}).  

(v) In a real accreting star, the accretion process itself continually pushes fluid elements through the crust.  A treatment that allows for this could show if this tends to reduce the strain built up by the increasing centrifugal forces.  

(vi) For simplicity, we followed \citet{Fattoyev_2018} and \citet{13} in considering the spin up of a spherical unstrained star.  The availability of a spherical unstrained background made the application of perturbation theory particularly straightforward.  More realistically, the pre-quake star would have, at the very least, an $(l=2, m=0)$ relaxed shape.  Closely related  to this, it would be good to model an isolated neutron star, being spun-down by, say, electromagnetic torques.  Such a star will surely have an $(l=2, m=0)$ relaxed shape.

To sum up, until now starquakes have only been modeled under the assumption of symmetric crust breaking. We found that non-axisymmetric starquakes are energetically allowed and hence, in future, it would be useful to build in-depth models for the asymmetric crust breaking. And, quite apart from its interest for continuous gravitational wave generation, non-axisymmetric shape changes may also have implications for the modelling of glitches themselves.  We therefore suggest that a combined glitch model, involving both crust fractures and superfluid unpinning, is worthy of exploration.  We defer further investigation of this interesting issue to future research.

\section*{Acknowledgements}

The authors thank Garvin Yim for his helpful comments on the final draft of this paper. YG acknowledges support from the Engineering and Physical Sciences Research Council (EPSRC) via grant No. EP/W524013/1.  DIJ acknowledges support from the Science and Technologies Funding Council (STFC) via grant No. ST/R00045X/1.

\section*{Data Availability}
This article did not use any data.



\bibliographystyle{mnras}
\bibliography{main} 



\newpage

\appendix

\section{Strain tensor}
\label{App1}
In this appendix, we will calculate the strain tensor

\begin{equation}
\label{b5}
\centering
\Sigma = 
\begin{bmatrix}
\Sigma_{rr} & \Sigma_{r\theta} & \Sigma_{r\phi}\\
\Sigma_{\theta r} & \Sigma_{\theta \theta} & \Sigma_{\theta \phi}\\
\Sigma_{\phi r} & \Sigma_{\phi \theta} & \Sigma_{\phi \phi}
\end{bmatrix}
= \frac{1}{2}(\xi^{AB}_{i;j} + \xi^{AB}_{j;i}).
\end{equation}
We are using the Kelvin mode displacement field vector for the $l=2$ and $m=2$ perturbation,

\begin{equation}
     \Vec{\xi_{22}^{AB}} = \alpha^{AB}_{22} r \left( 2  Y_{22}   \hat{e_r} + \frac{dY_{22}}{d\theta} \hat{e_\theta} - \frac{2}{\sin{\theta}}Y_{22} \hat{e_{\phi}}\right),
\end{equation}
where

\begin{equation}
    Y_{22}  = \frac{1}{4}\sqrt{\frac{15}{2\pi}}\sin^2{\theta}e^{i2\phi} 
\end{equation}

Expressions for the strain tensor components in spherical coordinates are taken from the book \cite{thorne2017modern} (Box 11.4). Given below is the complete strain tensor for the $l=2, m=2$ perturbation.
\begin{equation}
\label{b18}
\Sigma^{22} = \frac{\alpha^{AB}_{22}}{2}\sqrt{\frac{15}{2\pi}}
\begin{bmatrix}
  \sin^2{\theta}\cos{2\phi} &  \sin{\theta}\cos{\theta}\cos{2\phi} & - \sin{\theta}\sin{2\phi}\\
 \sin{\theta}\cos{\theta}\cos{2\phi} &  \cos^2{\theta}\cos{2\phi} &  - \cos{\theta}\sin{2\phi}\\
- \sin{\theta}\sin{2\phi} &  - \cos{\theta}\sin{2\phi} & - \cos{2\phi}
\end{bmatrix}
\end{equation}

In the similar way, one can calculate the strain tensor $\Sigma_{ij}^{20}$ for the axisymmetric perturbation $l=2, m=0$. For this case, we use the rotation displacement field (\ref{d1}). We get the following expressions for the components of the strain tensor $\Sigma_{ij}^{20}$ ,

\begin{equation}
    \Sigma_{rr}^{20} = \alpha^{AB}_{20}(3\cos^2{\theta}-1)(\frac{2}{R^2} +\frac{5r^2}{2R^2} -4),
\end{equation}

\begin{equation}
    \Sigma_{\theta \theta}^{20} = \alpha^{AB}_{20}\left[(\frac{5r^2}{2R^2} -4)(2-3\cos^2{\theta}) - \frac{r^2}{R^2} (3\cos^2{\theta}-1))\right],
\end{equation}

\begin{equation}
    \Sigma_{\phi \phi }^{20} = \alpha^{AB}_{20}\left[ - \frac{r^2}{R^2} (3\cos^2{\theta}-1) -  (\frac{5r^2}{2R^2} -4)\right],
\end{equation}

\begin{equation}
    \Sigma_{r \theta}^{20} = \Sigma_{ \theta r}^{20} = \alpha^{AB}_{20} \frac{6}{R^2}\sin{2\theta}(R^2 -r^2),
\end{equation}

\begin{equation}
    \Sigma_{ \theta \phi }^{20} = \Sigma_{ \phi \theta }^{20} = 0, 
\end{equation}

\begin{equation}
    \Sigma_{ r \phi }^{20} = \Sigma_{ \phi r }^{20} = 0. 
\end{equation}

\section{Alternative methodology to calculate the perturbed gravitational energy}
\label{App2}

As a check on our calculations for the $(l=m=2)$ perturbation in the gravitational potential energy of Section \ref{section 4}, we followed the method as given in \cite{SHAPIRO}. This method involves solving the Poisson's equation using Green's function in terms of spherical harmonics. We begin by solving the Poisson's equation 
\begin{equation}
\label{s1}
    \nabla^2 \Phi = 4 \pi G \rho,
\end{equation}
where $\rho$ is the matter density of the star and $\Phi$ is the gravitational potential of the star. We find the solution of (\ref{s1}) using Green's function
\begin{equation}
\label{s2}
    \Phi = - G \rho \int \frac{d^3 x'}{\lvert x - x' \rvert}.  
\end{equation}
In spherical coordinates we can expand $\frac{1}{\lvert x - x' \rvert}$ in terms of the spherical harmonics ($Y_l^m$):
\begin{equation}
\label{s3}
    \frac{1}{\lvert x - x' \rvert} = 4 \pi \sum_{l=0}^{\infty}\frac{1}{2l+1}\frac{r_<^l}{r_>^{l+1}}\sum_{m=-l}^{l}Y^m_l(\theta , \phi)Y^m_l(\theta' , \phi'),
\end{equation}
where $r_<$ is lesser of the two quantities $r$ and $r'$ and $r_>$ is the greater of the two quantities $r$ and $r'$. Inserting (\ref{s3}) into (\ref{s2}) we get
\begin{equation}
\label{s5}
    \Phi = -G \rho \int_V 4 \pi \sum_{l=0}^{\infty}\frac{1}{2l+1}\frac{r_<^l}{r_>^{l+1}}\sum_{m=-l}^{l}Y^m_l(\theta , \phi)Y^m_l(\theta' , \phi') dV,
\end{equation}
\begin{multline}
\label{s6}
    \Phi = -G \rho  4 \pi \int_0^{2\pi} \int_0^{ \pi}  \sum_{l=0}^{\infty}\frac{1}{2l+1} \left(\int_0^r \frac{(r')^{l+2} dr'}{r^{l+1}} \right. \\ \left. + \int_r^{R'} \frac{r^l dr'}{(r')^{l-1}}\right)\sum_{m=-l}^{l}Y^m_l(\theta , \phi)Y^m_l(\theta' , \phi') \sin{\theta'}d\theta' d\phi',
\end{multline}
where the domain of integration is over the volume of the triaxial ellipsoid. The polar equation of the surface of the ellipsoid is given as
\begin{equation}
\label{s4}
    \frac{\cos^2{\phi}\sin^2{\theta}}{a_1^2} +  \frac{\sin^2{\phi}\sin^2{\theta}}{a_2^2} +  \frac{\cos^2{\theta}}{a_3^2} = \frac{1}{R_s^2(\theta, \phi)},
\end{equation}
where $R_s(\theta, \phi)$ is radius of the triaxial ellipsoid and $a_1, a_2$ and $a_3$ are the radii along the $x, y$ and $z$ axis respectively related to a small dimensionless parameter as shown below:

\begin{equation}
\label{d11}
    a_1 = R(1 +\frac{\epsilon_{22}}{2}),
\end{equation}
\begin{equation}
\label{d12}
    a_2 = R(1 + \frac{\epsilon_{22}^2}{4}),
\end{equation}
\begin{equation}
\label{d13}
    a_3 = R(1 -\frac{\epsilon_{22}}{2}).
\end{equation}
where $\epsilon_{22}$ is the ellipticity parameter defined in equation (\ref{1}). The choice of radii $a_1, a_2$ and $a_3$ as shown above, conserves the volume of star up to the second order in $\epsilon_{22}$, which is essential for calculating the change in gravitational potential energy.   Solving (\ref{s6}) for $l=0$ and $l=2$ we get,
\begin{multline}
\label{d14}
     \Phi = \frac{2}{3}\pi G \rho [r^2 - 3 R^2] +  R^2 G\rho\frac{2 \pi}{15} \epsilon_{22}^2 + G \rho r^2 \frac{1}{4} (3 \cos^2{\theta} -1)\\(\frac{4 \pi \epsilon_{22}}{5} -\frac{11 \pi \epsilon_{22}^2}{10}) -\frac{ G \rho r^2 \pi \sin^2{\theta} \cos{2\phi}}{10} \epsilon_{22}.
\end{multline}
Next, we calculate the gravitational potential energy using
\begin{equation}
\label{s7}
    E_{\rm{grav}} = \frac{\rho}{2} \int_V \Phi \hspace{0.1 cm}d^3x,
\end{equation}
where V is the volume occupied by the triaxial ellipsoid. Inserting (\ref{d14}) into (\ref{s7}), we get
\begin{equation}
   E_{\rm{grav,s}} = - \frac{3 G M^2}{5 R}, \hspace{1cm} \delta E_{\rm{grav}} = \frac{ G M^2 }{25 R}\epsilon_{22}^2, \hspace{1cm} .
\end{equation}
$E_{\rm{grav,s}}$ is the leading order gravitational potential energy and $\delta E_{\rm{grav}}$ is the perturbed gravitational potential energy. This result agrees with the one we got earlier in Section \ref{section 4}.  We also used this method to calculate the change in gravitational potential energy for a pure $(l=2, m=0)$ perturbation, verifying successfully the result given in \cite{Baym_Pines_1971}, i.e.\ our equations (\ref{2.5a}) and (\ref{2.6}).

As (yet) another check, we also calculated the perturbed gravitational potential energy for ($l=2, m=2$) using the formalism of \cite{chandrasekhar1969ellipsoidal} and obtained the same result as above.


\bsp	
\label{lastpage}
\end{document}